\documentclass[aps,preprint,showpacs]{revtex4}
\usepackage{amsfonts}
\usepackage{amsmath}
\usepackage{amssymb}
\usepackage{graphicx}
\usepackage{epsfig}
\usepackage[english]{babel}
\usepackage{latexsym}
\usepackage{graphics,bm}
\usepackage{dcolumn}
\usepackage{rotating}
\usepackage[usenames]{color}
\usepackage{color}

\begin{document}

\title{Weakly interacting Bose gas in a random environment}

\author{G.M. Falco}
\affiliation{Institut f\"{u}r Theoretische Physik, Universit\"{a}t zu K\"{o}ln,
Z\"{u}lpicher Str. 77 D-50937 K\"{o}ln, Germany}
\author{T. Nattermann}
\affiliation{Institut f\"{u}r Theoretische Physik, Universit\"{a}t zu K\"{o}ln,
Z\"{u}lpicher Str. 77 D-50937 K\"{o}ln, Germany}
\author{V.L. Pokrovsky}
\affiliation{Dept. of Physics, Texas A\&M University, College Station, TX 77843-4242 and
Landau Institute for Theoretical Physics, Chernogolovka, Moscow District
142432, Russia}
\date{\today}
\begin{abstract}
Zero temperature properties of a dilute weakly interacting $d$-dimensional Bose gas in a random potential are
studied. We calculate geometrical and energetic characteristics of the localized state of a gas confined in
a large box or in a harmonic trap. Different regimes of the localized state are found depending on the
ratio of two characteristic length scales of the disorder, the Larkin length and the disorder correlation
length. Repulsing bosons confined in a large box with average density $n$ well below a critical value
$n_c$ are trapped in deep potential wells of extension much smaller than distance between them. Tunneling between 
these wells is exponentially small. The ground state of such a gas is a random singlet with no long-range phase correlation 
For $n>n_c$ repulsion between particles overcomes the disorder and the gas transits from the localized to a coherent 
superfluid state. The critical density $n_c$ is calculated in terms of the disorder parameters and the interaction strength. 
For atoms in traps four different regimes are found, only one of it is superfluid. The theory is extended to lower 
(1 and 2) dimensions. Its quantitative predictions can be checked in experiments with ultracold atomic gases
and other Bose-systems.
\end{abstract}

\maketitle


\section{Introduction}


\noindent Bose-Einstein condensation (BEC) is one of the key quantum phenomena
which shaped the physics of the 20th century. In BEC a macroscopic part of the
atoms is in the state with lowest energy. Well known examples are the
superfluid phase of $^{4}$He and laser cooled atoms in magnetic or optical
traps, which are a playground for the demonstration of generic quantum
phenomena like the appearance of quantized vortices, the Josephson effect etc.
\cite{Ketterle, Leggett, Dalfovo}.

A more complicated situation arises if these phenomena are studied in a random
environment like in porous media. Experimentally, superfluidity was shown to
persist in random media like Vycor or aerogel glass \cite{Reppy92} which
contain pores filling from $30\%$ up to $99\%$ of the material. Partial
coverage of the pores leads to a reduction of the transition temperature: the
transition temperature ranks from $2K$ for full-pore Vycor to about $5mK$ for
low coverage films in Vycor. Finally, for very low coverage superfluidity
corresponding to less than $1.5$ monolayers the superfluidity vanishes
completely even at $T=0$. One must presume that the helium is strongly
localized there. The remaining Helium atoms were considered to form an
\textquotedblright inert\textquotedblright\ layer which screens the random
potential once the density of the Helium atoms is again increased.

The experimental investigations on BEC in disordered environments concentrated
on critical behavior of the superfluid density and specific heat at decreasing
coverage of the porous media \cite{Reppy92}. More recently, Bose-Einstein
condensates have been studied in optical traps which superpose a harmonic and
a random potential \cite{Lye+05,Schulte+05,Fallani+07,Lugan+07,
Sanchez+07,Hulet+08}. Most of the papers concentrate on the possibility of
Anderson localization in effectively one-dimensional condensates. In contrast
to random potentials in crystals, the random speckle potential has a typical
correlations length of the order of $\mu m$ which is about of the same order
as the transverse extension of the atomic cloud. The longitudinal extension is
about $10$ to $100$ times larger {\cite{Lye+05}. In most of the experiments,
the expanding density profile after switching off the harmonic potential was
analyzed. \ } For weak random potential Lye et al. \cite{Lye+05} found
essentially the density profile corresponding to the Thomas-Fermi theory. For
larger values of the random potential these authors found a stripe-like
pattern of about $2\mu m$ spacing and damped dipole oscillations. For very
strong disorder the atoms are localized in the minima of the random potential.
Chen et al. \cite{Hulet+08} also studied \emph{in situ} and found
fragmentation of the BEC. In some papers the superposition of a periodic plus
a random potential was considered to enrich the number of possible phases
\cite{Schulte+05,Fallani+07}. Schulte et al. \cite{Schulte+05} found features
in the expanding condensate like pronounced fringes and changes in the axial
size of the central peak which they trace back to the random potential.

Theoretical investigations of Bose gases or fluids in a disordered environment
focused on three issues:

\noindent(i) The critical behavior at the normal to superfluid transition in
the vicinity of the transition temperature $T_{c}(n)$
\cite{Rasolt+84,Weichman+86,Weichman-Fisher86}, where $n$ denotes the boson
density. For fixed disorder strength $T_{c}(n)$ decreases with decreasing
boson density $n$ and eventually vanishes for $n\rightarrow n_{c}$.
\newline\noindent(ii) The description of zero-temperature quantum phase
transition from the superfluid to a non-superfluid (glassy) phase taking place
at the critical value $n_{c}$ of the boson density \cite{Ma+86, Giamarchi,
Fisher, Weichman07}. \newline\noindent(iii) The microscopic description of
the superfluid density (and the condensate) in a disordered environment in the
region of comparatively high average density of the gas $n\gg n_{c}$ and $T\ll
T_{c}(n)\,\,$ \cite{Lee-Gunn,Huang,Vinokur,Pitaevskii,Pelster,Graham}.

The problem of the Bose gas in the random environment attracted so much
attention because it connects two central ideas of the condensed matter
theory: the Bose condensation and localization. Especially the problem of the
interplay of the interaction and disorder is intriguing. So far this problem
was studied for Fermi gas \cite{altshuler-aronov, finkelstein}, but only in
the case of a weak disorder it could be done rigorously. Therefore the
delocalization induced by interaction remains beyond the frameworks of the
theory, though most of theorists are sure that it does exist. In the case of
the Bose gas the situation is even more involved since there is no Pauli
principle which delocalizes the ideal Fermi gas in 3 dimensions. Our study
shows that the interaction delocalizes the Bose gas at some critical
density.\noindent

Since in this paper we restrict ourselves mainly to zero temperature, below we give
a brief review only of the works mentioned in (ii) and
(iii). 
\newline\noindent(ii) The quantum phase transition, taking place at
zero temperature and $n=n_{c}$, exhibits in general a critical behavior
different from that observed at finite $T_{c}(n)$. This is because in the
quantum domain static and dynamic critical behavior are intertwined. This
quantum phase transition was considered by Ma, Halperin and Lee \cite{Ma+86}
who mapped the problem onto a $d+1$ dimensional classical $\phi^{4}$ theory
with disorder completely correlated in time. Their description gives rise to
new critical exponents which were calculated by $\epsilon$-expansion around
$d=4$ dimensions. This work has been criticized by Fisher et al. \cite{Fisher}.
One conclusion of the work of Fisher et al. \cite{Fisher} is that the
dynamic critical exponent $z$ should be equal to the dimension $d$ of the
system in all space dimensions. This result has been disputed recently \cite{Weichman07}. Indeed, quantum Monte Carlo studies show
$z\approx1.4$ for two-dimensional system, in discrepancy with the relation
$z=d$ \cite{Priya06}.

Some arguments of the work by Fisher et al. \cite{Fisher} rely on the
treatment of Giamarchi and Schulz \cite{Giamarchi} who found the superfluid to
insulator transition in one dimension. This transition happens at arbitrary
disorder by increasing the interaction to a sufficiently large value. The
fact that the transition into the non-superfluid
phase happens at increasing interaction seems paradoxical, but it can be
explained by the increase of quantum fluctuations of the phase. As we will
show, for weak interaction there is a second transition resulting from the
competition between the disorder and the interaction. We believe that namely
this transition has its counterparts in higher dimension and is observed in
experiments with cooled gases.

\noindent(iii) The fate of the inert layer and deeply localized (Lifshitz)
states was discussed by Lee and Gunn \cite{Lee-Gunn} without specific
predictions to be checked by experiments. More recently the reduction of the
superfluid and the condensate density deeply in the superfluid phase was
considered microscopically by Huang and Meng \cite{Huang}. These authors used
the Bogoliubov transformation and found a decrease of the superfluid density
and of the condensate if disorder is taken into account.
The corrections to both quantities are proportional to
$-\left(  {n_{c}}/{n}\right)  ^{1/2}$ in our terms. Thus this approach is restricted to
weak disorder and $n\gg n_{c}$. These results have been confirmed and
extended by other authors \cite{Pitaevskii, Vinokur, Pelster, Graham+Pelster,
Graham}. All these works used the approximation of  weak disorder. The
extrapolation of their results to the range of strong disorder made in
\cite{Huang} was illegitimate.

Zhou \cite{Zhou} argued that the $T=0$ transition to the superfluid phase is of the percolation type.
Lugan et al. \cite{Lugan+07} considered a BEC in a special  potential which is random in one and parabolic 
in the two remaining directions.
The disorder is bounded from below but not
from above. Using the numerical solution of the Gross-Pitaevskii equation as well as a mean field like approach they
identified three different regimes: the Lifshitz glass, the fragmented BEC and
the non-fragmented BEC. Because of the difference in the models (our potential includes randomness in all directions) a direct comparison with our results derived below is not possible, except for the one-dimensional case when the harmonic trap potential is negligible. Only in this case, and using our characteristics of the random potential, it can be shown that their approach gives  results  for the fragmented BEC comparable to ours. However we do not see a difference between the Lifshitz glass and  the fragmented BEC. Moreover, Lugan et al. \cite{Lugan+07}
  did not obtain important geometrical and energetic characteristics of the fragmented state vs. the average density of particles as we did based on a simple physical picture of the fragmentation.

More recent papers considered the dynamic consequences
of Anderson localization in BEC and do not directly overlap with our work
\cite{Paul+07,Lugan+07.2, Skipetrov08}. Using arguments similar to those used
in a published work by two of the authors \cite{Nattermann08}, Shklovskii
\cite{Shklovskii} independently and practically simultaneously arrived at the
same conclusions about the critical value of density at which the transition
between an insulating and the superfluid phase takes place, but he did not
give the description of the localized phase.

The aim of this paper is to develop an alternative approach to the quantum
phase transition starting from deeply localized state. We deliver a detailed
study of the deeply localized state of the Bose gas in a random potential.
Such a state appears at sufficiently low average density. We show, that in
this region the Bose-Einstein condensate decays into remote fragments of small sizes. We give a rather
transparent qualitative and quantitative geometrical
description of the localized state. We show that, at a critical density $n_{c}$ which we express in
terms of the disorder characteristics and interaction strength,
the increasing tunneling of particles between fragments leads to transition from the random singlet
state to the coherent superfluid. The localized state can exist in
several different regimes. These regimes as well as their quantitative characteristics strongly depend
on the ratio of two basic length scales characterizing the disorder: the Larkin
length $\mathcal{L}$ and the correlation length $b$. We extended our consideration to a practically
important situation of the Bose gas confined by a harmonic trap. In this case the gas forms a cloud
which size is determined by the energy minimization.
We demonstrate that all regimes which appear for the gas in the box appear also in the trap driven by
the number of particles and other parameters.
Our consideration is based on theory of deeply localized single-particle states in an uncorrelated
random potential given in seminal works by I. Lifshitz \cite{IMLifshitz}, Zittartz and Langer \cite{zittartz},
and Halperin and Lax \cite{halperin-lax}. Their predecessors Keldysh and Proshko \cite{Keldish}
and Kane \cite{Kane} studied electron density of state in a smooth random potential having in mind
semiconductors. These works were extended and detailized in References \cite{Bonch_a,Bonch_b, Andreev, Sh_Efros, cardy, John-Stephen, thirumalei}. 
In this work we propose a modest extension and simplification of their arguments to 
find some geometrical and physical characteristics of these state necessary for our purposes.
On the other hand, our theory can be considered as a generalization of their instanton-type theory
incorporating the self-consistent field of the interacting particles.

We consider the ground state of the system at zero temperature. It is not
always reachable since the system may be frozen in a metastable state. It is therefore important that nevertheless the
ground state occurs to be reachable
in a sufficiently wide range of parameters. We also demonstrate that
properties of the localized states of the Bose gas are rich enough and that they are available for the experimental study.

Our article is organized as follows. In the next section we characterize the random potential. In Section \ref{single} we
consider the single-particle deeply
localized states in 3 dimensions for the uncorrelated random potential obeying the Gaussian distribution. In Section \ref{Gas-in-box}
we consider how the dilute
Bose gas fills the deep fluctuation potential wells in this case. In Section \ref{transition} the transition from the random singlet
to the coherent superfluid state is
considered. Section \ref{corr-dis} extends all these results to the case of strongly correlated random potential. In Section \ref{gas-in-trap}
we consider the Bose gas subjected
to simultaneous action of a harmonic trap and random potential. Section \ref{low-dim}
represents extension of these results to lower dimensions 1 and 2. The
discussion of our results and conclusions are left for Section \ref{concl}.

\section{Description of the disorder\label{Descr-disorder}}
\noindent The disorder will be represented by a random potential $U\left(
\mathbf{x}\right)  $ with zero average at each point of the space and obeying
the Gaussian distribution:
\begin{align}
dW\left[  U\left(  \mathbf{x}\right)  ,dU\left(  \mathbf{x}\right)  \right]
&  =\exp\left(  -\frac{1}{2}\int U\left(  \mathbf{x}\right)  K^{-1}\left(
\mathbf{x,x}^{\prime}\right)  U\left(  \mathbf{x}^{\prime}\right)
d\mathbf{x}d\mathbf{x}^{\prime}\right) \label{gaussian}\\
&  \times\sqrt{\det K}~
{\displaystyle\prod\limits_{\mathbf{x}}}
\left(  \Delta\Omega/2\pi\right)  ^{1/2}dU\left(  \mathbf{x}\right),\nonumber
\end{align}
where $\Delta\Omega$ is the volume of an infinitesimal cell; $K\left(
\mathbf{x,x}^{\prime}\right)  $ is the correlation function or correlator of the random potential:%
\begin{equation}
K\left(  \mathbf{x,x}^{\prime}\right)  =\left\langle U\left(  \mathbf{x}%
\right)  U\left(  \mathbf{x}^{\prime}\right)  \right\rangle \label{correlator}%
\end{equation}
and $K^{-1}\left(  \mathbf{x,x}^{\prime}\right)  $ is the inverse correlator,
defined by equation:
\begin{equation}
\int K\left(  \mathbf{x,x}"\right)  K^{-1}\left(  \mathbf{x",x}^{\prime
}\right)  d\mathbf{x"}=\delta\left(  \mathbf{x-x}^{\prime}\right).
\label{inverse}
\end{equation}
\begin{figure}
\begin{center}
\includegraphics[width=0.7\linewidth]{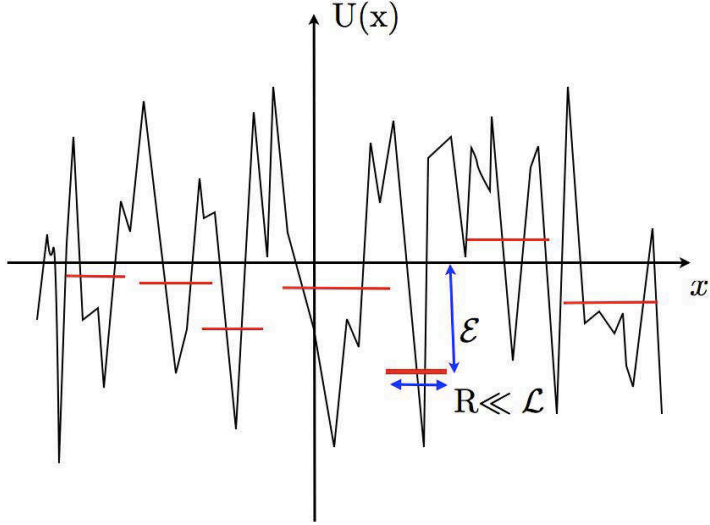}
\caption{\small{Uncorrelated potential, the characteristic length and energy scales are ${\cal L}$ and ${\cal E}$,
respectively. There is typically only one bound state in a single  potential well. }}
\label{Fig:uncorrelated_potential}
\end{center}
\end{figure}
The simplest random potential with zero correlation length has a $\delta-$like
correlator:
\begin{equation}
K_{0}\left(  \mathbf{x,x}^{\prime}\right)  =\left\langle U\left(
\mathbf{x}\right)  U\left(  \mathbf{x}^{\prime}\right)  \right\rangle
=\kappa^{2}\delta\left(  \mathbf{x-x}^{\prime}\right).  \label{white}
\end{equation}
We will call it uncorrelated random potential. The corresponding distribution
function is the product of independent distribution functions at each point of
space:
\begin{equation}
dW_{0}\left[  U\left(  \mathbf{x}\right)  ,dU\left(  \mathbf{x}\right)
\right]  =\exp\left(  -\frac{1}{2\kappa^{2}}\int U^{2}\left(  \mathbf{x}%
\right)  d\mathbf{x}\right)
{\displaystyle\prod\limits_{\mathbf{x}}}
\left[  \left(  \frac{\Delta\Omega}{2\pi\kappa^{2}}\right)  ^{1/2}dU\left(
\mathbf{x}\right)  \right]  \label{white-distribution}%
\end{equation}
More general Gaussian random potential has a finite correlation length $b$.
The simplest realization of the random Gaussian potential with the finite
correlation length is the Ornstein-Zernike correlator:
\begin{equation}
K_{OZ}\left(  \mathbf{x,x}^{\prime}\right)  =\frac{\kappa^{2}}{4\pi b^{2}%
}\frac{\exp\left(  -\frac{\left\vert \mathbf{x-x}^{\prime}\right\vert }%
{b}\right)  }{\left\vert \mathbf{x-x}^{\prime}\right\vert } \label{ornstein}%
\end{equation}
The corresponding probability distribution reads:%
\begin{align}
dW_{OZ}\left[  U\left(  \mathbf{x}\right)  ,dU\left(  \mathbf{x}\right)
\right]   &  =\exp\left(  -\frac{1}{2\kappa^{2}}\int\left[  U^{2}\left(
\mathbf{x}\right)  +b^{2}\left(  \nabla U\left(  \mathbf{x}\right)  \right)
^{2}\right]  d\mathbf{x}\right) \label{corr-distribution}\\
&
{\displaystyle\prod\limits_{\mathbf{q}}}
\left[  \left(  \frac{\left(  1+b^{2}q^{2}\right)  }{2\pi\kappa^{2}}%
\Delta\Omega_{\mathbf{q}}\right)  ^{1/2}d\tilde{U}\left(  \mathbf{q}\right)
\right]  ,\nonumber
\end{align}
where $\tilde{U}\left(  \mathbf{q}\right)  $ is the Fourier-transformation of
$U\left(  \mathbf{x}\right)  $ and $\Delta\Omega_{\mathbf{q}}$ is the element
of volume in the momentum space. At $b=0$ Eq. (\ref{corr-distribution}) turns
into Eq. (\ref{white-distribution}). The OZ distribution has a characteristic
energy scale $U_{0}=\kappa/b^{3/2}$.

The disorder correlator may be regular at coincident coordinates
$\mathbf{x=x}^{\prime}$. Then it has the following form:
\begin{equation}
K\left(  \mathbf{x,x}^{\prime}\right)  =\left\langle U^{2}\right\rangle
{h}\left(  \frac{\left\vert \mathbf{x-x}^{\prime}\right\vert }{b}\right),
\label{finite corrlength}%
\end{equation}
where $\left\langle U^{2}\right\rangle \equiv\left\langle U^{2}\left(
\mathbf{x}\right)  \right\rangle $ is the average quadratic fluctuation of the
random field at a point. The function $h\left(  u\right)  $ which determines
the shape of the correlation function is normalized by the condition $h\left(
0\right)  =1$. In this case the Fourier-component of the inverse correlator is
a growing function of the wave-vector. As a consequence, $K^{-1}\left(
\mathbf{x,x}^{\prime}\right)  $ is not well defined. Instead we can use the
probability in the momentum space:%
\begin{equation}
dW\left[  \tilde{U}\left(  \mathbf{q}\right)  ,d\tilde{U}\left(
\mathbf{q}\right)  \right]  =\exp\left(  -\int\frac{\tilde{U}\left(
\mathbf{q}\right)  \tilde{U}\left(  -\mathbf{q}\right)  }{2\tilde{K}\left(
\mathbf{q}\right)  }d\mathbf{q}\right)
{\displaystyle\prod\limits_{\mathbf{q}}}
\left(  \frac{\Delta\Omega_{\mathbf{q}}}{2\pi\tilde{K}\left(  \mathbf{q}%
\right)  }\right)  ^{1/2}d\tilde{U}\left(  \mathbf{q}\right)  \label{momentum}%
\end{equation}
Simple examples of non-singular correlators are the Lorenz distribution with
$h_{L}\left(  u\right)  =\left(  u^{2}+1\right)  ^{-1}$ and $\tilde{K}
_{L}\left(  \mathbf{q}\right)  =2\pi^{2}\left\langle U^{2}\right\rangle
b^{2}q^{-1}\exp\left(  -bq\right)  $ and the Gaussian distribution with
$h_{G}\left(  u\right)  =$ exp$\left(  -u^{2}/2\right)  $ and $\tilde{K}
_{G}\left(  \mathbf{q}\right)  =\left(  2\pi\right)  ^{3/2}\left\langle
U^{2}\right\rangle b^{3}\exp\left(  -b^{2}q^{2}/2\right)$. In all these
cases the scale of energy is established by $U_{0}=\sqrt{\left\langle
U^{2}\right\rangle }$ and the scale of length is $b$.

\section{Single-particle levels in an uncorrelated random
potential\label{single}}

In this section we study the properties of single-particle states in the
uncorrelated random potential defined by Eqs. (\ref{white},
\ref{white-distribution}). This is a modest extension of the well-known works
by I.M. Lifshitz \cite{IMLifshitz}, Zittartz and Langer \cite{zittartz} and
Halperin and Lax \cite{halperin-lax}. Some simplifications allow us to find
more detailed information about the distribution functions of the energy,
sizes, distances between the states and the tunneling amplitudes between them.
We are interested in a statistical description of the spectrum and wave
functions of Schr\"{o}dinger equation in the random potential $U\left(
\mathbf{x}\right)  $:%
\begin{equation}
\frac{\hslash^{2}}{2m}\nabla^{2}\psi+\left(  E-U\left(  \mathbf{x}\right)
\right)  \psi=0 \label{schroedinger}%
\end{equation}

Its energy levels $E\left[  U\left(  \mathbf{x}\right)  \right]  $ in a finite
volume are functionals of the potential $U\left(  \mathbf{x}\right)  $.  The
only characteristic of the random potential $\kappa$ together with world
constants $\hslash$ and $m$ establishes the scale of length:
\begin{equation}
\mathcal{L}=\frac{\hslash^{4}}{m^{2}\kappa^{2}} \label{larkin}
\end{equation}
which will be called Larkin length in analogy with the scale found in the
Larkin's work \cite{Larkin-70} for an elastic medium in a random field. We
will show that ${\cal L}$
sets the scale of the extension of the deeply localized states with large
by modulus negative energy. For delocalized states with large positive energy
the same value with precision of numerical factor is the mean free path. The
scale of energy $\mathcal{E}$ can be obtained from the Larkin length in an
obvious way:
\begin{equation}
\mathcal{E}=\frac{\hslash^{2}}{m\mathcal{L}^{2}}=\frac{m^{3}\kappa^{4}%
}{\hslash^{6}} \label{L-energy}%
\end{equation}
Further we will work in a rather rough approximation similar to that used by
Larkin and Ovchinnikov \cite{LO} and Imry and Ma \cite{imry-ma}. However, we
start with a rigorous statement of the problem which gives a clue for our
further estimates. The most easily calculable value is the density of state
$\nu\left(  E,\Omega\right)  $ which can be written as a path integral (see
the cited works \cite{IMLifshitz},\cite{zittartz},\cite{halperin-lax}):%
\begin{equation}
\nu\left(  E,\Omega\right)  =\frac{1}{\Omega}\int\delta\left(  E-E\left[  U\left(
\mathbf{x}\right)  \right]  \right)  dW\left[  U\left(  \mathbf{x}\right)
,\Omega\right]  , \label{DOS}
\end{equation}
where $E\left[  U\left(  \mathbf{x}\right)  \right]  $ is the spectrum of
eigenvalues of the Schr\"{o}dinger equation (\ref{schroedinger}) in the volume
$\Omega$.

In a large 3d volume the states with energy $E\gg\mathcal{E}$ are delocalized,
whereas the states with negative energy sufficiently large by modulus $E<0$
and $\left\vert E\right\vert \gg\mathcal{E}$ are strongly localized. The
threshold of localization is a positive energy of the order of $\mathcal{E}$.
Note that this is correct only for 3d systems. As it was conjectured in the
work by Abrahams et al. \cite{abrahams}, all single-particle states in 1 and
2d systems are localized.
Near the threshold the wave function has very
complicated fractal structure \cite{mirlin}, \cite{feigel}. In the interval
between $\mathcal{E}$ and $-\mathcal{E}$ the transition from fractal to strongly
localized states evolves. The latter are supported by rare
fluctuations of the random potential, which form a potential well sufficiently
deep to have the negative energy $E$ as its only bound state. The problem of the deep
random levels was considered by already mentioned authors
\cite{IMLifshitz}, \cite{zittartz}, \cite{halperin-lax} about 40 years ago. We
reproduce some of their results and extend them to find the probability
distribution of the levels with energy less than some $E$ ($E<0;\left\vert
E\right\vert \gg\mathcal{E}$), the distances between such states and the
tunneling amplitude between them.

As it is clearly seen from equations (\ref{gaussian},\ref{DOS}), the main
exponential factor in the density of state can be found by minimization of the
integral $\int_{\Omega}U^{2}\left(  \mathbf{x}\right)  d\Omega$ at a fixed
value of the energy level $E\left[  U\left(  \mathbf{x}\right)  \right]  $,
which is a functional of the random potential $U\left(  \mathbf{x}\right)  $.
The latter can be determined as a minimum of the energy over the wave
function:%
\begin{equation}
E\left[  U\left(  \mathbf{x}\right)  \right]  =\min_{\psi\left(
\mathbf{x}\right)  }E\left[  U\left(  \mathbf{x}\right)  ,\psi\left(
\mathbf{x}\right)  \right]  =\min_{\psi\left(  \mathbf{x}\right)  }\int\left[
\frac{\hslash^{2}}{2m}\left\vert \nabla\psi\left(  \mathbf{x}\right)
\right\vert ^{2}+U\left(  \mathbf{x}\right)  \left\vert \psi\left(
\mathbf{x}\right)  \right\vert ^{2}\right]  d\Omega\label{energy-functional}%
\end{equation}
Thus, we need to minimize a functional:
\begin{equation}
F\left[  U\left(  \mathbf{x}\right)  ,\psi\left(  \mathbf{x}\right)  \right]
=\int U^{2}\left(  \mathbf{x}\right)  d\Omega-\lambda\int\left[  \frac
{\hslash^{2}}{2m}\left\vert \nabla\psi\left(  \mathbf{x}\right)  \right\vert
^{2}+U\left(  \mathbf{x}\right)  \left\vert \psi\left(  \mathbf{x}\right)
\right\vert ^{2}\right]  d\Omega\label{functional}%
\end{equation}
over $\psi\left(  \mathbf{x}\right)  $ and $U\left(  \mathbf{x}\right)  $.
Here $\lambda$ is a Lagrangian factor. The minimization over $\psi\left(
\mathbf{x}\right)  $ leads to Schr\"{o}dinger equation (\ref{schroedinger}),
whereas the minimization over $U\left(  \mathbf{x}\right)  $ results in a
relation between $U\left(  \mathbf{x}\right)  $ and $\psi\left(
\mathbf{x}\right)  $:%
\begin{equation}
U\left(  \mathbf{x}\right)  =\lambda\left\vert \psi\left(  \mathbf{x}\right)
\right\vert ^{2} \label{U-psi-square}%
\end{equation}
Thus, equation (\ref{schroedinger}) turns into the Ginzburg-Landau equation.
For our purpose the most important consequence of the relationship
(\ref{U-psi-square}) is that the fluctuation potential well $U\left(
\mathbf{x}\right)  $ has the same characteristic linear size $R$ as the wave
function $\psi\left(  \mathbf{x}\right)  $. It is clear that the maximum
probability requires the bound state with the fixed energy $E$ to be the only
bound state in the potential well. Otherwise, at the same energy, we need a
deeper well, i.e. larger $U^{2}\left(  \mathbf{x}\right)  $. For the same
reason the fluctuation well must have the spherical shape. Let the radius of
the well is $R$. Then the depth of the well can be estimated as $U_{\min}
\sim-\frac{\hslash^{2}}{mR^{2}}$ and the energy level in it differs by a
factor of the order of 1/2: $E\sim-\frac{\hslash^{2}}{2mR^{2}}$. The
exponential factor in the density of state reads:
\begin{equation}
\exp\left(  -\frac{\frac{4\pi}{3}R^{3}U^{2}}{2\kappa^{2}}\right)  =\exp\left(
-\frac{\mathcal{L}}{R}\right)  =\exp\left[  -\left(  \frac{\left\vert
E\right\vert }{\mathcal{E}}\right)  ^{1/2}\right]  , \label{exp}%
\end{equation}
where we redefined the Larkin length $\mathcal{L}$ incorporating the factor
$2\pi/3$ in it. Here and henceforth we perform calculations for 3d systems.
The results for other dimensions will be derived later. The result (\ref{exp})
is valid provided the number in the exponent is large, i.e. $R\ll\mathcal{L}$
and $\left\vert E\right\vert \gg\mathcal{E}$.

\begin{figure}
\begin{center}
\includegraphics[width=0.7\linewidth]{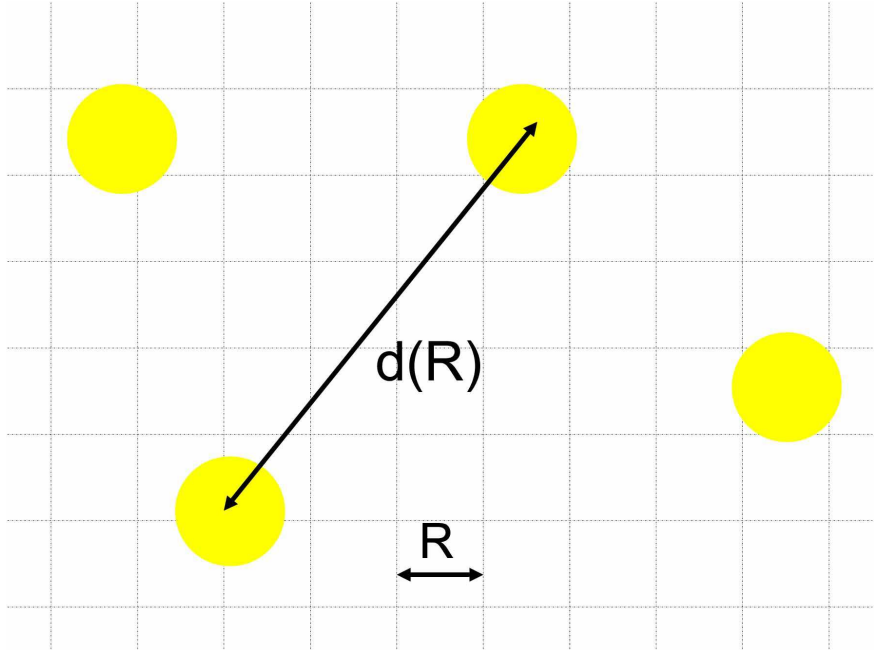}
\caption{\small{Deeply localized states. For low densities of  bosons the distance
between deeply localized states of radius $R\ll{\cal L}$ (or smaller)
are separated by distances $d(R)\gg {\cal L}$. }}
\label{Fragments}
\end{center}
\end{figure}

Let us consider the probability or the part of volume $q\left(  R\right)  $
occupied by the wells with the radius less than $R$ or the energy less than
$E=-\frac{\hslash^{2}}{2mR^{2}}$. It is obvious that $q\left(  R\right)  $
contains an exponential factor (\ref{exp}) and some preexponent. Since
$q\left(  R\right)  $ is a dimensionless value, the preexponent must be a
function of the dimensionless ratio $\frac{\mathcal{L}}{R}$:%
\begin{equation}
q\left(  R\right)  =f\left(  \frac{\mathcal{L}}{R}\right)  \exp\left(
-\frac{\mathcal{L}}{R}\right)  \label{probability}%
\end{equation}
The function $f\left(  x\right)  $ must be much slower function of its
argument than the exponent. Most naturally it is a power function $f\left(
x\right)  \sim x^{\alpha}$ with a critical exponent $\alpha$.
It can be extracted from the work by J. Cardy \cite{cardy}: $\alpha =1$ for
the uncorrelated disorder. We will see that it is inessential for further conclusions.
By knowledge of $q\left( R\right)  $ we can calculate the number $n_{w}\left(  R\right)  $
of the wells with the radius less than $R$ per unit volume, i.e. the density of such wells \cite{d-w}.
In order to do that the unit volume must be divided into the cells of the
volume $R^{3}$. Each cell can contain or do not contain the fluctuation, but
fluctuations with the centers approaching each to other at a distance less
than $R$ must be considered as one asymmetric potential well. The number of
such cells in unit volume is $R^{-3}$. Thus:%
\begin{equation}
n_{w}\left(  R\right)  =R^{-3}q\left(  R\right)  =R^{-3}f\left(
\frac{\mathcal{L}}{R}\right)  \exp\left(  -\frac{\mathcal{L}}{R}\right)
\label{dens-wells}%
\end{equation}
The average distance $d\left(  R\right)  $ between the wells with the radius
less than $R$ reads:%
\begin{equation}
d\left(  R\right)  =\left(  n_{w}\left(  R\right)  \right)  ^{-1/3}%
=Rf^{1/3}\exp\left(  \frac{\mathcal{L}}{3R}\right)  \label{distance-R}%
\end{equation}
This equation shows that the distance between the wells is significantly
larger than the size of the wells. The situation is depicted in Fig.~\ref{Fragments}.
Finally we find the tunneling factor $t\left(  R\right)  $ for two typical
wells with the energy levels of the same order or the radius $R$ of the same
order of magnitude. It is given by the semiclassical expression
$t\left(R\right)  =\exp\left(  -\frac{1}{\hslash}\int\left\vert p\right\vert
dl\right)  $, where the integral path connects two wells. By the order of
magnitude $\left\vert p\right\vert \sim\sqrt{2m\left\vert E\right\vert }%
\sim\hslash/R$ and the length of the path of integration is $\sim d\left( 
R\right)  $. Thus, $\frac{1}{\hslash}\int\left\vert p\right\vert dl\sim
d/R=f^{1/3}\exp\left(  \frac{\mathcal{L}}{3R}\right)  $. Finally we find:
\begin{equation}
t\left(  R\right)  =\exp\left[  -f^{1/3}\exp\left(  \frac{\mathcal{L}}%
{3R}\right)  \right]  \label{tunneling-R}%
\end{equation}
At $R\sim\mathcal{L}/3$ or $E\sim-9\mathcal{E}$, the distances between the
optimal potential wells become of the same order of magnitude as their size
$R$. Simultaneously the tunneling amplitude between the wells becomes of the
order of 1. The potential wells percolate and tunneling is not small, but the
states still are not propagating due to the Anderson
localization \cite{anderson}.


\section{Bose gas in a large box with an uncorrelated random
potential\label{Gas-in-box}}


In the ground state of an ideal Bose gas in a large box with the Gaussian
random potential all particles are located at the deepest fluctuation level.
In the box of the cubic shape with the side $L$ the deepest level which occurs
with probability of the order of 1 has the radius $R$ determined by equation:
$L^{3}n_{w}\left(  R\right)  =1$, i.e. $R\sim\frac{\mathcal{L}}{3\ln\left(
L/\mathcal{L}\right)  }$. Note that the prefactor introduces a correction to
the denominator of the order of $\ln\left(  \ln\frac{L}{\mathcal{L}}\right)  $
which can be neglected. The corresponding energy is $E\sim-9\mathcal{E}\left(
\ln\frac{L}{\mathcal{L}}\right)  ^{2}$. Such a state is highly non-ergodic
since the location and the depth of the deepest level strongly depends on a
specific realization of the disordered potential. Therefore the average energy
per particle and other properties averaged over the ensemble has nothing in
common with the properties of a specific sample. Even an infinitely small
interaction immediately changes the situation: the system becomes ergodic in
the thermodynamic limit, i.e. when first the size of the system grows to
infinity and then the interaction goes to zero. For example, the energy per
particle in sufficiently large volume coincides with its average over the
ensemble. The reason of such a sharp change is that, at any small but finite
interaction, the particles can not more fill one well since their repulsion
finally overcomes the attraction to the potential well. They will be
redistributed over multiple wells. Since the distribution of wells in
different parts of sufficiently large volume passes all possible random
configurations with proper ensemble probabilities, the ergodicity is
established. In this section we find how the interacting particles eventually
fills deep single-particle levels simultaneously changing their size and shape
by their self-consistent field. Similar idea was considered in the work by
Lugan \textit{et al}.\cite{Lugan+07.2}. However, in this work the
consideration was done at fixed chemical potential instead of fixed number of
particles. Besides, they did not perform geometrical analysis and calculation
of the tunneling amplitude for single-particle states. Therefore, their
results do not contain our explicit expressions for the geometrical structure
and tunneling of the localized states as well as the analysis of the
transition line.

In a real experiment the Bose gas may be quenched in a metastable state
depending on the cooling rate and other non-thermodynamic factors. This is
what M.P.A. Fisher \textit{et al.} \cite{Fisher} call the Bose glass. Such a
state is also possible in the case of weakly repulsive Bose gas. It looks plausible
that the Bose glass whose metastable state has energy close to that of the ground state
has also close to the ground state statistical characteristics, but this question needs
more thorough study. However, as it will be demonstrated later, in the case of cooled 
alcali atoms the tunneling amplitude still remains large enough to ensure the relaxation to the
equilibrium state in 10$^{-3}\div$10$^{-2}s$. Our further estimates relate to
the real ground state.

The weakly repulsive Bose gas in a random potential is described by the
well-known Hamiltonian:
\begin{equation}
H=\sum_{\mathbf{p}}\frac{\mathbf{p}^{2}}{2m}a_{\mathbf{p}}^{\dag}%
a_{\mathbf{p}}+\frac{g}{2}\int\left(  \psi^{\dag}\psi\right)  ^{2}%
d\mathbf{x+}\int U\left(  \mathbf{x}\right)  \psi^{\dag}\psi d\mathbf{x,}
\label{WIHamiltonian}%
\end{equation}
where $\psi\left(  \mathbf{x}\right)  =\Omega^{-1/2}\sum_{\mathbf{p}
}a_{\mathbf{p}}\exp\left(  i\mathbf{px/}\hslash\right)  $ is the secondary
quantized wave function; the positive coupling constant $g$ is associated with
the scattering amplitude $a$ by the relationship\cite{LLQM}:
\begin{equation}
g=\frac{4\pi\hslash^{2}a}{m} \label{g-a}%
\end{equation}
As in the Bogoliubov's theory \cite{bogoliubov} and its extension by
Belyaev \cite{belyaev}, we assume that the gas criterion $na^{3}\ll1$ is
satisfied. Here $n=N/\Omega$ is the average particle density and $N$ is their
total number. The purpose of the following consideration is to find how, with
the growth of the number of particles, they eventually fill the fluctuation
potential wells. Implicitly our considerations takes in account the change of
the optimal potential wells due to the interaction.

Let $N$ particles with the average density of particles $n$ fill all potential
wells with the radii less than $R$ in the ground state. The average number of
particles per well is
\begin{equation}
\mathcal{N}\left(  R\right)  =\frac{n}{n_{w}\left(  R\right)  }
\label{number -per-well}%
\end{equation}
The local density inside the well of the linear size of $R$ is $n_{p}
(R)=\frac{3{\cal{N}}\left(  R\right)  }{4\pi R^{3}}$. The gain of energy per particle
due to the random potential is $E\left(  R\right)  =-\frac{\hslash^{2}}{2mR^{2}}$;
the repulsion energy due to interaction is equal to $gn_{p}\left(  R\right)
=\frac{3\hslash^{2}{\cal{N}}\left(  R\right)  a}{mR^{3}}$, where we used Eq.~(\ref{g-a})
and the well-known relation for an effective potential field
induced by a gas of scatterers \cite{LLQM}. Minimizing the total energy per
particle $E_{tot}\left(  R\right)  =-\frac{\hslash^{2}}{2mR^{2}}
+\frac{3\hslash^{2}\mathcal{N}\left(  R\right)  a}{mR^{3}}$ over $R$ and
employing Eqs. (\ref{dens-wells}) and (\ref{number -per-well}), we find the
value of $R$ corresponding to the minimum of energy at fixed $n$ with the
logarithmic precision:%
\begin{equation}
R\left(  n\right)  =\frac{\mathcal{L}}{\ln\frac{n_{c}}{n}}, \label{R-n}%
\end{equation}
where the critical density $n_{c}$ is defined as follows:%
\begin{equation}
n_{c}=\left(  3\mathcal{L}^{2}a\right)  ^{-1} \label{n-crit}%
\end{equation}
The factor $f$ in Eq. (\ref{dens-wells}) as well as the next
approximation to the solution (\ref{R-n}) leads to the corrections of the type
$\ln\left(  \ln\frac{n_{c}}{n}\right)  $ which can be neglected. The distances
between the filled wells according to Eq. (\ref{distance-R}) are:%
\begin{equation}
d\left(  n\right)  =\frac{\mathcal{L}}{\ln\frac{n_{c}}{n}}\left(  \frac{n_{c}%
}{n}\right)  ^{1/3} \label{r-n}%
\end{equation}
This distance strongly exceeds the average size of the potential well
(\ref{R-n}) at $n\ll n_{c}$. At the same condition the chemical potential of
atoms can be estimated as:%
\begin{equation}
\mu\left(  n\right)  =-\frac{\hslash^{2}}{2mR^{2}\left(  n\right)  }%
=-\frac{\mathcal{E}}{2}\left(  \ln\frac{n_{c}}{n}\right)  ^{2}
\label{chem-pot}%
\end{equation}
The tunneling amplitude $t\left(  n\right)  $ between two wells separated by a
typical distance $r\left(  n\right)  $ can be found by employing the single
particle result (\ref{tunneling-R}):
\begin{equation}
t\left(  n\right)  =\exp\left[  -\left(  \frac{n_{c}}{n}\right)
^{1/3}\right] . \label{tunneling-n}%
\end{equation}

Thus, the Bose gas at $n\ll n_{c}$ is fragmented into a multiple clusters of
small size $R\left(  n\right)  $ separated by much larger distances $d\left(
n\right)  $ and containing about $\mathcal{L}/\left[  3a\left(  \ln\frac
{n_{c}}{n}\right)  ^{3}\right]  $ particles each. The amplitude of tunneling
between the wells is exponentially small, so that the number of particles in
each cluster is well defined. The phase is therefore completely uncertain.
Such a state is a singlet with inhomogeneously distributed particles, a random
singlet: the ground state is non-degenerate.

The previous approach is valid if $\mathcal{N}\left(  R\right)  \gg1$. From
Eqs. (\ref{dens-wells}, \ref{number -per-well}, \ref{n-crit}) we find
$\mathcal{N}\left(  R\right)  =\mathcal{L}/\left[  3a\left(  \ln\frac{n_{c}
}{n}\right)  ^{3}\right]  $. Thus, the disorder must be weak enough
to satisfy the constraint $\mathcal{L\gg}3a$. But it is not sufficient.
Additionally the average density $n$ must be large enough. Namely, it must
satisfy strong inequalities $\exp\left[  -\left(  \frac{\mathcal{L}}
{3a}\right)  ^{1/3}\right]  n_{c}\ll n\ll n_{c}$. At smaller $n$, only few
particles appear in a potential well. At $n\ll\exp\left[  -\left(
\frac{\mathcal{L}}{3a}\right)  ^{1/3}\right]  n_{c}$, the size of filled
potential wells approaches $36a$. Most of the filled wells contain 2
particles, but only a small part of the total number of wells of this size
(per unit volume) $n_{w}\left(  36a\right)  $ is filled, approximately
$0.5\times10^{-4}a^{-3}\exp\left(  -\frac{\mathcal{L}}{36a}\right)  $. The
consistency of our approximation requires that the gas parameter remains small inside an optimal
potential well. The density inside the well is equal to $n_{c}$.
Therefore, the interaction is weak inside the well if $n_{c}
a^{3}=\frac{a}{3\mathcal{L}}\ll1$.

The compressibility $\partial n/\partial \mu=\frac{n}{\cal E}\ln\left(\frac{n_c}{n}\right)$ is finite as
expected for the Bose glass phase \cite{Fisher}.
In the absence of interaction the probability that two single-particle eigenstates localized in different
potential wells have the same energy is  zero.
Thus the zero temperature conductivity vanishes in the thermodynamic limit. If the interaction is switched on, we can
consider its effect as splitting of the energy levels by an amount of the order $gn_p$. For $n\ll n_c$ the wave
functions are still localized and the conductivity
at zero temperature vanishes. Thus, the Bose glass phase is insulating. At finite temperatures however the
energy gap between  different localized states can be bridged by
thermal activation leading to the Mott variable range hopping (VRH) \cite{Efros}. Below we apply the VRH argument to our situation.
The tunneling probability between two wells,
each of extension $R$, separated by the distance $L$, is $\mid t(L,R)\mid=\exp(-2L/R)$ (see analogous derivation in section 3).
Together with the activation factor it gives the
hopping probability between the two wells equal to
\begin{equation}\label{e:VRH}
   P(T)\sim e^{-2L/R}e^{-\Delta E/T},
\end{equation}
where $\Delta E$ is the energy difference between the two distant wells. There is with high
probability at least one localized state in the interval
of energy $\Delta E$ provided the condition $\Delta E \nu(E) L^3\gtrsim 1$ is satisfied.
Employing the derived dependencies
$E\approx -\hbar^2/(2mR^2)$, $R={\cal L}/\ln(n_c/n)$ and $\nu(E) = {\cal E}^{-1}R^{-3}\exp{(-{\cal L}/R)}$
and maximizing the probability
(\ref{e:VRH}) over $L$, we find the hopping conductivity $\sigma(T)$
\begin{equation}\label{e:VRH2}
   \sigma(T)\sim e^{-C\left[{\cal E}n_c/(Tn)\right]^{1/4}},
\end{equation}
where C is a constant or slow function of $\ln{n_c/n}$. In a similar way it is
possible to calculate the non-linear field-dependent conductivity.


\section{Transition region\label{transition}}


Though our results are valid only for $n\ll n_{c}$, it follows from them that at $n\sim n_{c}$
the overlapping of different occupied wells becomes large and
the tunneling amplitudes reach the value of the order of 1.
Therefore, the
phase coherence in different wells grows until at some critical value of
density which we roughly identify with $n_{c}$ the global coherence is
established. Thus, at $n\approx n_{c}$, the quantum phase transition from
localized singlet to the superfluid state proceeds. As usual, the order
parameter is the average phase factor $\left\langle \exp\left(  i\varphi
\right)  \right\rangle $. The superfluid state bears clear traces of disorder:
its superfluid density is inhomogeneous at characteristic scale $\mathcal{L}$.
It can be treated as a disordered superfluid.
An alternative argument goes as follows. The condition $n\approx n_c$ can be rewritten in the
form ${\cal L}\approx\xi$ where $\xi=(an)^{-1/2}$ denotes the
superfluid healing length.
The latter describes the range over which the
superfluid order parameter, when  changed by a local perturbation   (e.g. by a wall or in
the center of a vortex), regains its bulk value. Since ${\cal L}$
denotes the scale at which the disorder becomes relevant, the condition  $\xi \gg {\cal L}$
simply means that
the superfluid order parameter  cannot adapt to the  rapid variation of the disorder which
happens at the scale ${\cal L}$ and superfluidity is destroyed. In the opposite
case
$n\gg n_c$, i.e. $\xi\ll {\cal L}$ this is not longer the case and superfluidity is  restored.

At $n\gg n_{c}$ the energy of the repulsion $gn$ becomes much larger than the
characteristic disorder energy $\mathcal{E}$. Then the gas becomes almost
homogeneous with precision of small parameter $\mathcal{E}/gn\sim n_{c}/n$. In
this case the approach by Huang and Meng \cite{Huang} based on the homogeneous
zero-approximation ground state and refined by Giorgini \textit{et al.
}\cite{Pitaevskii}\textit{\ }and by Lopatin and Vinokur \cite{Vinokur} is
justified. They demonstrated that the superfluid density $n_{s}$ is not equal
to the total density as it is without disorder, but remains close to the total
density. {Thus, the superfluidity definitely strives in the limit of large
density}.
We conjecture that $n_{s}$ vanishes at the same quantum phase
transition point $n=n_{c}$ at which the coherence disappears.

The small corrections found in the cited work \cite{Huang} can be estimated by
the order of magnitude without long calculations. Indeed, the disorder
Hamiltonian reads:
\begin{equation}
H_{dis}=\sum_{\mathbf{p,q}}U_{\mathbf{q}}a_{\mathbf{p}}^{\dag}a_{\mathbf{p+q}%
}, \label{disorder}
\end{equation}
where $U_{\mathbf{q}}$ is Fourier transform of the random field $U\left(
\mathbf{x}\right)  $. It produces the change of energy in the second order of
the perturbation theory. The change of energy per particle can be estimated as
follows:
\begin{equation}
\Delta\varepsilon=\int\frac{\left\langle U_{\mathbf{q}}^{2}\right\rangle
}{\varepsilon_{\mathbf{q}}}\frac{d^{3}q}{\left(  2\pi\hslash\right)  ^{3}}
\label{perturb}
\end{equation}
The delta-like correlation function of the random potential (\ref{white})
implies that $\left\langle U_{\mathbf{q}}^{2}\right\rangle =\kappa^{2}$ for
any $\mathbf{q}$. The excitation energy $\varepsilon_{\mathbf{q}}$ can be
estimated as $\sqrt{\frac{gn}{m}}q$ in the effective range of integration
$q<\sqrt{mgn}$ after subtraction of the contribution of disorder to the energy
of the normal state. Collecting all these factors, we find:
\begin{equation}
\Delta\varepsilon\approx\frac{\kappa^{2}m^{3/2}\left(  gn\right)  ^{1/2}}
{4\pi^{2}\hslash^{3}}=\frac{\kappa^{2}\left(  man\right)  ^{1/2}}{2\sqrt{\pi
}\hslash^{2}} \label{huang-corr}%
\end{equation}
With precision of a numerical coefficient 0.85 this result coincides with that
obtained by Huang and Meng (Equation (7) of their work \cite{Huang} gives the
value of energy per unit volume; it must be divided by $n$ to find the energy
of the ground state per particle). In terms of the Larkin length and energy
and critical density introduced earlier (see Eqs. (\ref{larkin},
\ref{L-energy}, \ref{n-crit})) the correction to the energy of the ground
state per particle can be expressed as
\begin{equation}\label{eq:Deltaepsilon}
\Delta\varepsilon\approx\frac
{\sqrt{\pi}}{3}\mathcal{E}\left(  \frac{n}{n_{c}}\right)  ^{1/2}\sim\frac{\hbar^2}{m\xi^2}\frac{\xi}{\cal L}.
\end{equation}
The change
of superfluid and condensate densities can be found from their initial values
(one exactly and another approximately equal to the total density $n$) by
multiplication to a small factor proportional to $\Delta\varepsilon
/\varepsilon_{0}\sim \xi/{\cal L}$, where $\varepsilon_{0}=\frac{gn^{}}{2}=\frac{2\pi
\hslash^{2}an^{}}{m}\sim\frac{\hbar^2}{m\xi^2}$.

In discussing critical phenomena  it is crucial to determine the critical (marginal)
dimensions. In order to do that we need some results which will be derived later.
We will show that the critical density at the transition between the superfluid and the Bose glass phase in $d$ dimensions reads
\begin{equation}\label{eq:n_c(d)}
n_c\approx \frac{1}{a^{d-2}{\cal L}_d^2} ,\quad \text{i.e.} \quad \xi_d\approx {\cal L}_d,
\end{equation}
where ${\cal L}_d=(\hbar^2/m\kappa)^{2/(4-d)}$ and $\xi_d\sim (a^{d-2}n)^{-1/2}$  denote the $d$-dimensional Larkin and healing length, respectively. For strong disorder and
strong interaction the natural limits of the validity of equation (\ref{eq:n_c(d)})
seem to be $\xi_d\gg  n^{-1/d}$ and ${\cal L}_d\gg n^{-1/d}$, i.e. the healing length and the Larkin length must be much larger than the mean spacing between the bosons.

For weak disorder, the Larkin length diverges at $d\to 4$. Hence in dimensions $d>4$ disorder has to overcome a threshold value to destroy superfluidity even in the limit of weak interaction.
Thus $d=4$ is an upper critical dimension of the problem as has been discussed already
in \cite{Fisher}. A lower critical dimension is $d=1$. This can be seen from a calculation of (\ref{eq:Deltaepsilon}) in one dimension which is logarithmically infrared divergent reflecting
strong quantum fluctuation. Giamarchi and Schulz \cite{Giamarchi} in a seminal paper found a quantum phase transition in 1d Bose gas in disordered environment from a superfluid to an insulating phase
for increasing interaction parameter $K\approx2/3$ (compare Appendix). Since then this transition has been considered
as the genuine  transition between these two phases (see e.g. \cite{Fisher}). However, at this transition $\xi_1 n\approx 0.25$ strongly violating the gas condition $\xi_1n\gg 1$. Thus,
we conclude that the transition found in \cite{Giamarchi} is not the one-dimensional equivalent of the transition taking place at $n\approx n_c$. The phase portrait obtained in this work is
reproduced in Fig.~\ref{Figure8}.
It follows from this figure that there are two different phase transitions at small disorder: one at large value of dimensionless coupling constant $K$ and another at small $K$. The large part
of the phase diagram beyond a small vicinity of the critical point
$K=2/3$, $\kappa=0$  was obtained by a speculation rather than a rigorous treatment. Therefore, the question about the existence
of two different quantum phase transition remains open.

Critical behavior near the quantum transition was studied by Halperin, Lee and Ma\cite{Ma+86} and by M.P.A. Fisher \textit{et al.}\cite{Fisher}. In the latter work the scaling
consideration was developed and even the critical exponents were found. However, we are not sure that they are valid for our problem since, according to previous remark, they are
associated with the strong-coupling critical point. Secondly, even if there is only one critical point, the marginal dimensionality is 1+1 and it is very doubtful that exact
critical exponent can be found in the case 3+1.

\section{Correlated disorder\label{corr-dis}}

In this section we consider properties of the weakly interacting Bose gas in a big box with a correlated random potential. The properties of correlated random potential
were already described in the second half of Section
\ref{Descr-disorder} (see the text related to Eqs. (\ref{finite corrlength}
-\ref{momentum})). Here, following the line of consideration developed in
Sections \ref{single}-\ref{transition}, we study first the size and
distribution of single-particle states and tunneling between them and how they
are modified by interaction. We start with the single-particle states.

\subsection{Single-particle states in a correlated random potential}

Keldysh and Proshko \cite{Keldish}, and Kane \cite{Kane} were the first to find
the electron density of states in a semiclassical random potential. They proved that,
in contrast to the uncorrelated disorder for which $\nu\left(  E\right)
\sim\exp\left[  -\left(  E/\mathcal{E}\right)  ^{1/2}\right]  $, the density
of state in a correlated disorder is a Gaussian function of energy $\nu\left(
E\right)  \sim\exp\left[  -c\left(  E/U_{0}\right)  ^{2}\right]  $, 
where $c$ is a numerical constant. Independently
but later the same result was found in \cite{Bonch_a,Bonch_b,Andreev}. 
Shklovskii and Efros \cite{Sh_Efros} derived the same result employing the instanton
approach.    
John and Stephen \cite{John-Stephen} found the dependence of the preexponential
factor on energy. Their result was confirmed by Thirumalai \cite{thirumalei} by a 
different method.  Below we rederive their result and find new important
characteristics of optimal fluctuation wells in the case when the correlation
length is large enough. It is intuitively clear that, at $b\ll\mathcal{L}$,
the density of state and other characteristics of the spectrum and optional
states only slightly differ from their values for uncorrelated random
potential. At $b\sim\mathcal{L}$ numerical coefficients will be different from
those for uncorrelated disorder, but with this reservation still our
semiquantitative description is valid. Therefore, the most interesting is the situation with $b\gg\mathcal{L}$. Such a disorder we call strongly correlated.
In what follows we assume the strongly correlated disorder . The Larkin length
for the correlated disorder reads:
\begin{equation}
\mathcal{L=}\frac{3\hslash^{4}}{4\pi m^{2}U_{0}^{2}b^{3}} \label{larkin-corr}
\end{equation}
It is convenient to introduce another characteristic length $B=\left(
\frac{3}{4\pi}\right)  ^{1/4}\left(  \frac{\hbar^2}{mU_{0}}\right)  ^{1/2}$.
For strongly correlated disorder $B\ll b$. The Larkin length is associated with
$B$ by the following relation:
\begin{equation}
\frac{\mathcal{L}}{b}=\left(  \frac{B}{b}\right)  ^{4} \label{larkin-B}
\end{equation}
\begin{figure}
\includegraphics[width=0.7\linewidth]{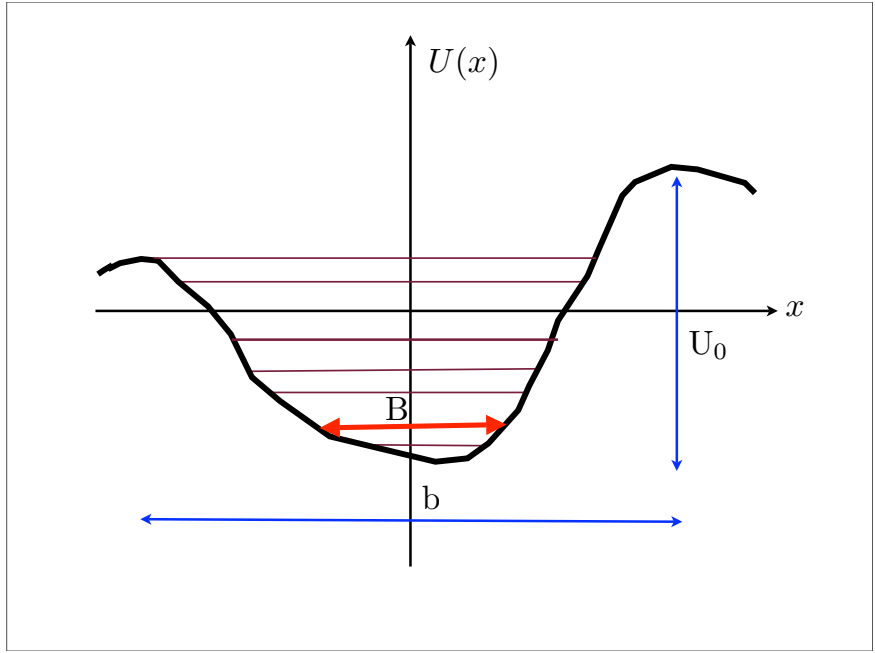}
\caption{\small{Long range correlated potential. There are typically many bound states per well. }}
\label{Figure3}
\end{figure}
To find the exponent in the density of deeply localized states it is necessary
to minimize the functional $\int U\left(  \mathbf{x}\right)  K^{-1}\left(
\mathbf{x,x}^{\prime}\right)  U\left(  \mathbf{x}^{\prime}\right)
d\mathbf{x}d\mathbf{x}^{\prime}$ at fixed energy of the quantum state in the
potential $U\left(  \mathbf{x}\right)  $. The condition of minimum is reduced
to a following relationship between the optimal fluctuation potential
$U\left(  \mathbf{x}\right)  $ and the wave function \ $\psi\left(
\mathbf{x}\right)$:
\begin{equation}
U\left(  \mathbf{x}\right)  =\lambda\int K\left(  \mathbf{x,x}^{\prime
}\right)  \left\vert \psi\left(  \mathbf{x}^{\prime}\right)  \right\vert
^{2}d\mathbf{x}^{\prime}, \label{U-psi2-corr}
\end{equation}
analogues to Eq. (\ref{U-psi-square}) for the uncorrelated disorder. This
equation shows that the characteristic size of $U\left(  \mathbf{x}\right)  $
is always equal to $b$. Indeed, it is correct if the characteristic size of the
wave function less or equal to $b$. In the opposite case the characteristic
size of the potential would be much larger than $b$ and, respectively, the
probability of such configuration would be much smaller. Deep levels have
negative energy $E$, much larger by modulus than the energy scale of the
random potential $U_{0}$. Since the optimal fluctuation potential well must
have the level $E$, its depth is not less than $\left\vert E\right\vert $. The
fundamental difference between the optimal potential wells in the cases of
uncorrelated and strongly correlated disorder is that the former contains only
one level, whereas the latter contains many levels as shown in
Fig~\ref{Figure3}. Indeed, the number of
levels in the optimal well is not less than $\frac{1}{6\pi^2}\left(
\sqrt{2m\left\vert E\right\vert }b/\hslash\right)  ^{3}\geqslant\frac{1}{6\pi^2
}\left(  \sqrt{2mU_{0}}b/\hslash\right)  ^{3}\sim\left(  \frac{b}{\mathcal{L}   
}\right)  ^{3/4}=\left(  \frac{b}{B}\right)  ^{3}\gg1$.
Therefore, the level
$E$ is located close to the bottom of the optimal potential well. Its depth
with high precision is equal to $\left\vert E\right\vert $ and its shape is
determined by the function $h\left(  r/b\right)  $ from Eq.
(\ref{finite corrlength}). If the correlation function is non-singular, then
$U\left(  \mathbf{x}\right)  =E~h\left(  r/b\right)  $. The exponent in
Eq.~(\ref{gaussian}) can be calculated exactly leading to the result for the
probability $q\left(  E\right)  $ to find the level with energy less than $E$
(with precision of a prefactor):
\begin{equation}
q\left(  E\right)  =\exp\left(  -\frac{E^{2}}{2\left\langle U^{2}\right\rangle
}\right)  \label{probability-corr}
\end{equation}
In a more accurate treatment one needs to take in account that the level $E$
can be not the ground state in the optimal well, but an excited state. Since
even excited states are close to the bottom, the spectrum is the same as for
a spherically symmetric quantum oscillator with the frequency $\omega=\left(
\left\vert E\right\vert h''\left(  0\right)  /mb^{2}\right)  ^{1/2}$.
Therefore, the depth $V$ of the well is not exactly equal to $\left\vert
E\right\vert $. It depends on what excited state occupies the particle:
\begin{equation}
V_{n_{1},n_{2},n_{3}}=\left\vert E\right\vert +\left[  \left(  n_{1}
+n_{2}+n_{3}\right)  +3/2\right]  \hslash\omega\label{excited}
\end{equation}
The second term in the depth of well determined by this equation is much less
than the first one. After summation over all $n_{i}$ $\left(  i=1,2,3\right)
$ from 0 to infinity we find the corrected expression for the probability
$q\left(  E\right)  $:
\begin{equation}
q\left(  E\right)  =\exp\left(  -\frac{E^{2}+3\left\vert E\right\vert
\hslash\omega}{2\left\langle U^{2}\right\rangle }\right)  \left[
1-\exp\left(  -\frac{\left\vert E\right\vert \hslash\omega}{\left\langle
U^{2}\right\rangle }\right)  \right]  ^{-3} \label{summation}
\end{equation}
The ratio $\left\vert E\right\vert \hslash\omega/\left\langle U^{2}
\right\rangle $ which appears in Eq. (\ref{summation}) by the order of
magnitude is equal to the product of two ratios $\left(  \left\vert
E\right\vert /U_{0}\right)  ^{3/2}\times\left(  {{B}}/b\right)$.
The first factor is large, the second one is small and their product varies
from small to large values always remaining larger than $\left(
B/b\right) $. Nevertheless, a strong enhancement of
probability up to a factor $\left(  b/{B}\right)  ^{3}$ in
comparison to the initial expression (\ref{probability-corr}) is reached at
decreasing modulus of energy. Summarizing, we conclude that the participation
of excited states may change a comparatively slow-varying prefactor leaving
the principal exponent (\ref{probability}) unchanged.

Next we analyze the case of a singular correlator using the example of the Ornstein-Zernike
distribution. The relationship (\ref{U-psi2-corr}) is valid for singular
correlators implying that the radial size of the optimal potential well is
equal to the correlation length $b$. We use the functional
(\ref{corr-distribution}) to estimate the exponent in the probability
$q\left(  E\right)  $. The two terms in the exponent (\ref{corr-distribution})
proportional to $U^{2}$ and $b^{2}\left(  \nabla U\right)  ^{2}$ are equal.
With the precision of a numerical constant the potential is equal to
$E\left(b/r\right)  e^{-r/b}$. The integration leads to $q\left(  E\right)
=\exp\left(  -cE^{2}/U_{0}^{2}\right)  $ in accordance with the earlier
results \cite{John-Stephen}, \cite{thirumalei}. Further we incorporate the constant
$c$ into the definition of $U_{0}$ and write the equation for the probability
in all cases as:
\begin{equation}
q\left(  E\right)  =\exp\left(  -E^{2}/2U_{0}^{2}\right)
\label{probability-general}
\end{equation}
For simplicity we put the prefactor to be equal to 1.

The density of potential wells containing deep levels lower than $E$ for
strongly correlated random potential reads:
\begin{equation}
n_{w}\left(  E,b\right)  =b^{-3}\exp\left(  -E^{2}/2U_{0}^{2}\right)
\label{n-w-corr}
\end{equation}
and the average distance between such wells is
\begin{equation}
d\left(  E,b\right)  =b\exp\left(  E^{2}/6U_{0}^{2}\right)
\label{distance-corr}
\end{equation}
The tunneling amplitude between such wells can be found in semiclassical
approximation:
\begin{align}
t\left(  E,b\right)   &  =\exp\left(  -\frac{\sqrt{2m\left\vert E\right\vert
}d}{\hslash}\right)  =\label{tunneling-corr}\\
&  \exp\left[  -\frac{b}{B}\exp\left(  E^{2}/6U_{0}^{2}\right)  \right]
\nonumber
\end{align}
It is always very small, but the direct tunneling between wells does not play
an important role since the particles remain almost classical even near the
mobility edge in contrast to the uncorrelated disorder. According to the Shklovsky
estimate \cite{Shklovskii}, the percolation of wells happens at $E=-0.9U_{0}$.
The characteristic De-Broglie wavelength $\hslash/\sqrt{2m\left\vert
E\right\vert }$ at this energy is much less than $b$ for strongly correlated
disorder. Therefore, the particles propagate according to classical mechanics.
The mobility edge energy is negative and not far from the percolation threshold.

\subsection{Weakly repulsive Bose gas in strongly correlated disordered potential}

In this section we consider the Bose gas with the average density $n$. The gas partly fills all potential wells with the depth less than $E$. Let the
particles fill a typical well up to the radius $R<b$. If the correlator is
regular at $\mathbf{x}=\mathbf{x}^{\prime}$, the
following equation gives
a rough estimate of energy per particle:
\begin{equation}
\mu\left(  E,R;b,n\right)  =E\left(  1-\frac{R^{2}}{2b^{2}}\right)
+\frac{3\hslash^{2}na}{m}\left(  \frac{b}{R}\right)  ^{3}\exp\left(
\frac{E^{2}}{2U_{0}^{2}}\right)  \label{energy-per-particle-corr}
\end{equation}
The first term in this expression is the gain of energy due to the random
potential, the second term is the energy of interaction. The quadratic term in
the brackets corresponds to the oscillator-like potential at small $R$.
Minimizing $\mu\left(  E,b,R\right)  $ over $E$ and the ratio $R/b$, one
arrives at following results:
\begin{align}
E  &  \approx-U_{0}\sqrt{2\ln\frac{n_{c}}{n}};\label{E-n-corr}\\
\frac{R}{b}  &  \approx\left(  \ln\frac{n_{c}}{n}\right)  ^{-1/2}.
\label{ratio}
\end{align}
The critical density $n_{c}$ is defined as follows:
\begin{equation}
n_{c}\approx\frac{\sqrt{3}}{4}\frac{mU_{0}}{\hslash^{2}a}\sim\frac{1}{B^{2}a}
\label{crit-density-corr}
\end{equation}
All results are valid at $n\ll n_{c}$. According to Eq.~(\ref{ratio}), the
ratio $\frac{R}{b}$ is small. This fact justifies the approximation of quadratic
potential in Eq.~(\ref{energy-per-particle-corr}). It allows to estimate the 
characteristic momentum of particles or the width of their distribution over momentum:
\begin{equation}
\Delta p=m\omega R=m\omega b\left(  \ln\frac{n_{c}}{n}\right)  ^{-1/2}
\label{momentum-corr}
\end{equation}

The chemical potential is readily derived from Eqs.
(\ref{energy-per-particle-corr}-\ref{crit-density-corr}):
\[
\mu\left(  b,n\right)  =-U_{0}\sqrt{2\ln\frac{n_{c}}{n}}\left[  1-\frac{9}
{4}\left(  \ln\frac{n_{c}}{n}\right)  ^{-1}\right]
\]
Two filled wells are separated by a typical distance $d\left(  b,n\right)  $
much larger than the size of the well:
\begin{equation}
d\left(  b,n\right)  =b\left(  \frac{n_{c}}{n}\right)  ^{1/3}
\label{distance-n-corr}
\end{equation}
With this value of $d$ it is ready to estimate the tunneling coefficient:
\begin{equation}
t\left(  b.n\right)  \approx\exp\left[  -\frac{b}{B}\left(  \frac{n_{c}}
{n}\right)  ^{1/3}\right]  \label{tunneling-n-corr}%
\end{equation}
The number of particles in each typical well can be estimated as
$\mathcal{N}=\frac{n}{n_{w}\left(  E\right)  }=n_{c}b^{3}\approx\frac{b^{3}
}{B^{2}a}$. It is large if $b\gtrsim a$.

These results conclude the description of localized states for strongly
correlated disorder. At energies between $-U_{0}$ and $U_{0}$ the motion of
particle is strongly disordered, but it becomes more and more free when energy
approaches and exceeds $U_{0}$. At $E\gg U_{0}$, the random potential can be
considered as a perturbation. The quantum phase transition from the localized
random singlet state to a superfluid proceeds at $n=n_{c}$. Note that the
critical density in the presence of strongly correlated disorder does not
depend on the correlation length $b$.

\section{Bose gas in a trap with disordered potential\label{gas-in-trap}}

In this section we consider deeply localized states in a harmonic trap
supplied with a Gaussian random potential. The Hamiltonian of the system has
an additional term: the potential energy of Bose particles in the trap:
\begin{equation}
H_{trap}=\int V_{trap}\left(  \mathbf{x}\right)  \psi^{\dag}\left(
\mathbf{x}\right)  \psi\left(  \mathbf{x}\right)  d\mathbf{x,}
\label{trap energy}
\end{equation}
where the harmonic potential of the trap generally has a form:
\begin{equation}
V_{trap}\left(  \mathbf{x}\right)  =\frac{m}{2}\left(  \omega_{x}^{2}
x^{2}+\omega_{y}^{2}y^{2}+\omega_{z}^{2}z^{2}\right)  \label{trap potential}
\end{equation}
In this section we consider only the isotropic trap with $\omega_{x}
=\omega_{y}=\omega_{z}=\omega$. Strongly anisotropic traps can be considered
as a system with reduced dimension 2 or 1 and will be considered in the next
section. A new scale of length associated with the trap is the well-known
oscillator length $\ell=\sqrt{\hslash/\left(  m\omega\right)  }$.

A principal difference between the Bose gas in a box and in a trap is that in
the latter case the gas forms a cloud whose size is determined by energy
minimization at a fixed number of particles, whereas in the former \ case the
size of the box and, therefore, its average density $n$ is fixed.
Nevertheless, we will see that all phase states of the Bose gas in the box
appear when this gas is placed into a harmonic trap, but their appearance is
regulated by the total number of particles in the gas $N$ instead of their
average density.

\subsection{Trap with uncorrelated disorder}

Four competing parts of a Bose-particle energy are: kinetic energy, the
confining potential energy of the trap, the repulsion from other particles and
the energy of the random potential. Two of them, the interaction with the trap
and the random potential tend to confine and localize the particle. Let the
gas form a single cloud of the radius $R$. The rough estimate of the two
confining energy contributions can be done as follows \cite{Nattermann08}%
\begin{equation}
V_{trap}=\frac{m\omega^{2}R^{2}}{2}=\frac{\hslash^{2}}{2m}\frac{R^{2}}%
{\ell^{4}} \label{trap-R}
\end{equation}
The disorder energy for the same cloud can be estimated as%
\begin{equation}
V_{dis}=-\frac{\kappa}{\left(  \frac{4\pi R^{3}}{3}\right)  ^{1/2}}%
\approx-\frac{\hslash^{2}}{2m}\left(  \frac{\mathcal{L}}{R}\right)  ^{3/2}
\label{disorder-R}
\end{equation}
The latter estimate can be justified as follows. The cloud chooses such a
position that the random potential energy is negative. A characteristic
fluctuation of energy in a volume $\Omega$ is
\[
\frac{1}{\Omega}\left\langle \left(  \int_{\Omega}U\left(  \mathbf{x}\right)  d\mathbf{x}%
\right)  ^{2}\right\rangle ^{1/2}=\frac{1}{\Omega}\left[
{\displaystyle\iint\limits_{\Omega}}
K\left(  \mathbf{x,x}^{\prime}\right)  d\mathbf{x}d\mathbf{x}^{\prime}\right]^{1/2} 
= \frac{\kappa}{\sqrt{\Omega}}
\]
Substituting into the last expression $\Omega=\frac{4}{3}\pi R^{3}$, we arrive
at Eq. (\ref{disorder-R}). Comparing the two confining contributions, we
conclude that, at $\mathcal{L}\gg\ell$, the influence of disorder is
negligible. The remaining two terms of energy: kinetic energy
\begin{equation}
K\sim\hslash^{2}/\left(  2mR^{2}\right)  \label{kinetic}%
\end{equation}
and the repulsion energy per particle:%
\begin{equation}
V_{int}=gn=\frac{\hslash^{2}}{m}\frac{3Na}{R^{3}} \label{repulsion}%
\end{equation}
are deconfining. The kinetic energy dominates at $R\gg3Na$. In the opposite
case $R\ll3Na$ the repulsion energy dominates. Below we analyze several
limiting cases.\newline
\begin{figure}
\begin{center}
\includegraphics[width=0.7\linewidth]{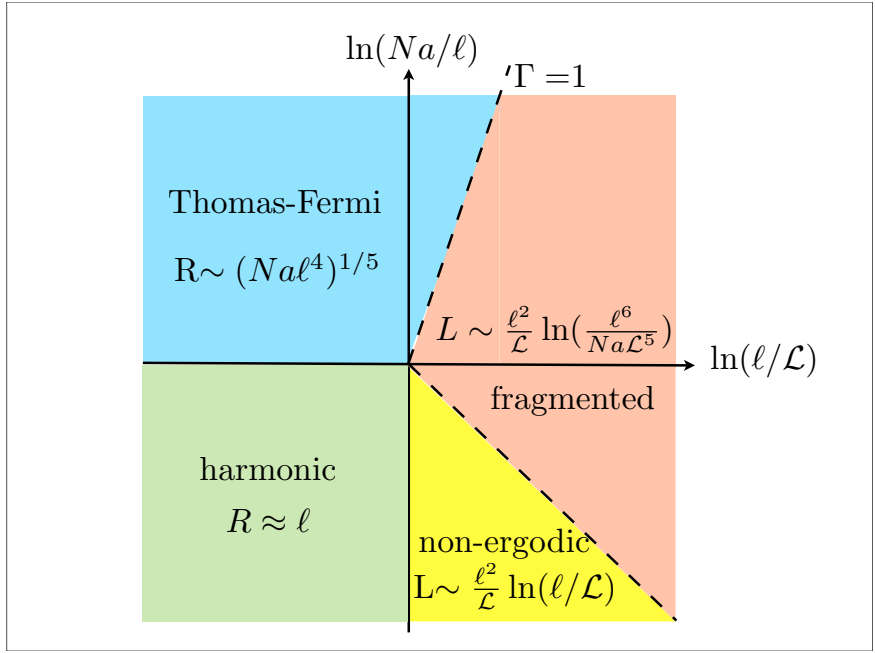}
\caption{\small{Regime diagram of atoms in three dimensional traps: uncorrelated disorder.  $R$
denotes the size of the single existing atomic cloud. $L$ is the size of the cloud of fragments.}}
\label{Fig:regime_diagram_uncorrelated}
\end{center}
\end{figure}

\noindent1 \textbf{Weak disorder}:\textit{ }$\mathcal{L}\gg\ell$. Depending
on the relative value of the interaction one should distinguish two different situations:

\noindent1a. \textit{Weak interaction: }$3Na\ll\ell$. In this case the
interaction can be neglected. Minimizing too remaining terms, the kinetic
energy and energy of the trap, we find $R=\ell$. Physically it means that all
particles are condensed at the oscillator ground state.

\noindent1b. \textit{Strong interaction: }$3Na\gg\ell$. Neglecting kinetic
energy and minimizing remaining energy of traps plus the repulsion energy, one
finds the result known as Thomas-Fermi approximation \cite{Stringari}: $R=\left(
\frac{9}{2}Na\ell^{4}\right)  ^{1/5}$.

More interesting is the case of strong disorder.

\noindent2. \textbf{Strong disorder:}\textit{ }$\mathcal{L}\ll\ell$. Again we
consider several different limiting cases depending on relative strength of
disorder, interaction and the trap potential.

\noindent2a. \textit{Weak interaction: }$3Na\ll\mathcal{L}\ll\ell$. In this
range of variables the non-ergodic phase is realized. Since interaction is
negligible, the particles find a random potential well with the deepest level
and fall into it. Let such a well can be found at a distance $\sim L$ from the
trap center. Its depth typically is about $9\mathcal{E}\ln^{2}\left(
L/\mathcal{L}\right)  $. This gain of energy must be not less than the loss of
the trap energy $m\omega^{2}L^{2}/2$. A typical value of $L$ appears when both
this energies have the same order of magnitude. Then $L\approx6\sqrt{2}\left(
\ell^{2}/\mathcal{L}\right)  \ln\left(  \ell/\mathcal{L}\right)  $. A typical
size of the well is $R\approx\mathcal{L}/\left(  6\ln\left(  \ell
/\mathcal{L}\right)  \right)  $.

\noindent2b. \textit{Moderate interaction: }$\mathcal{L}\ll3Na\ll\ell$. In
this case the ergodicity is restored. Our experience with the gas in a box
prompts that the gas cloud is split into fragments each occupying a random
potential well from very small size till some size $R$ depending on $N$. From
Eq. (\ref{chem-pot}) the typical disorder energy per particle is
$\mu=-\mathcal{E}\left(  \ln\frac{n_{c}}{n}\right)  ^{2}$. It becomes equal to
the trap energy at a distance $L\sim\left(  \ell^{2}/\mathcal{L}\right)
{\ln\Gamma}$ (see the definition of the parameter $\Gamma$ below in Eq.~(\ref{Gamma})).
Therefore, the average density is $n\sim N\mathcal{L}^{3}/\ell^{6}$. The ratio
$n_c/n$ plays the decisive role: the state of the Bose gas is fragmented and strongly
localized when it is large; the transition to delocalized superfluid state
proceeds when this ratio becomes 1. With precision of a numerical factor the
ratio $n_{c}/n$ is equal to a new dimensional parameter:%
\begin{equation}
\Gamma=\frac{\ell^{6}}{3Na\mathcal{L}^{5}} \label{Gamma}%
\end{equation}
Strongly localized fragmented state is realized at $\Gamma\gg1$. The
transition proceeds at $\Gamma=1$. The superfluid phase emerges at $\Gamma<1$.
The regime diagram is shown in Fig. \ref{Fig:regime_diagram_uncorrelated}.\cite{regime} 
Note the counter-intuitive dependence of the cloud size on the number of particles: the cloud
slightly contracts with increasing number of particles. It happens because the
number of particles in each fragment increases more rapidly with the average
density than the number of fragments.

\subsection{Trap with strongly correlated disorder}

Following the line of development accepted in the previous subsection, we
start with the case of weak disorder.

\noindent1. \textbf{Weak disorder: }$\frac{\ell}{b+\ell}U_{0}\ll\hslash\omega
$ or ${B}\gg\ell$. The value in the left-hand side of this interaction is 
the characteristic variation of the
random potential on the minimal of two linear sizes: oscillator length $\ell$
or correlation length $b$. In this case the random potential only slightly
changes the state of the gas in the absence of disorder (see previous
subsection). In what follows we assume for simplicity that $\ell\geq b$.

\noindent2. \textbf{Strong disorder: }$U_{0}\gg\hslash\omega$ or $B\ll\ell$.
As in the previous subsection we consider different relative strength of the interaction and disorder.

\noindent2a. \textit{Weak interaction: }$3Na\ll\frac{b^3}{B^{2}}
\ll\ell$. The interaction is negligible. Therefore, all particles find the
deepest well in the trap and fall into it. The typical distance from the
center of the trap to the deepest well is $L=\frac{\ell^2}{B}\left(
\ln\frac{\ell^2}{b B}\right)^{1/4}$
and its typical depth is $E=-U_{0}\sqrt{6\ln\frac{L}{b}}$. In
this range of parameters the system is non-ergodic.

\begin{figure}
\begin{center}
\includegraphics[width=0.7\linewidth]{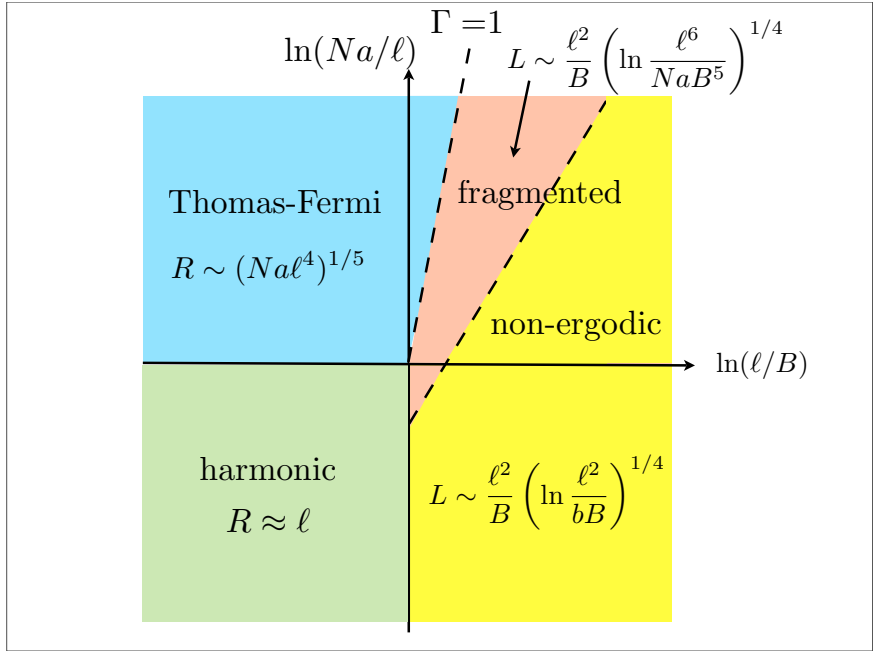}
\caption{\small{Regime diagram of atoms in three dimensional traps: correlated disorder.
$R$ denotes the size of the single atomic cloud, $L$ denotes the size of the fragmented state.}}
\label{Figure5}
\end{center}
\end{figure}

\noindent2b. \textit{Moderate interaction: }$\frac{b^3}{B^{2}}
\ll3Na\ll\ell$. In this case the particles are distributed among multiple
potential wells. The system is ergodic. The average depth of the well is
$E\left(  N\right)  =-U_{0}\sqrt{2\ln\Gamma}$, where
\begin{equation}
\Gamma=\frac{U_{0}^{5/2}}{\hslash^{2}Nam^{1/2}\omega^{3}}=\frac{\ell^{6}
}{NaB^{5}} \label{Gamma-corr}
\end{equation}
is the dimensionless parameter analogues to that introduced in the previous
subsection. It represents the ratio $n_{c}/n$. Until $\Gamma$ is large the
ground state of the Bose gas is deeply localized random singlet. The
transition to the superfluid state proceeds at $\Gamma\sim1$. \ The total
linear size of the cloud is $L=2^{3/4}\frac{\ell^{2}}{B}\left(  \ln
\Gamma\right)  ^{1/4}$. Inside a fluctuation potential which has a typical
size $b$ particles occupy a smaller sphere of the radius $R\sim\left(
Nab^{2}B^{2}\right)  ^{1/5}$. The phase state (regime) of the system depends 
on three dimensionless parameters: $B/b$, $3Na/\ell$ and $\ell/b$. The regime diagram is
shown on Fig. \ref{Figure5}.

\section{Lower dimensions\label{low-dim}}
In the experiments with cooled gases the trap is mostly realized as a rotation
ellipsoid with a large aspect ratio. Therefore, and also for comparison with
exact 1d solution \cite{Giamarchi} it is reasonable to extend our theory for lower
dimensions $d=1,2$.

\subsection{Uncorrelated disorder}

We start with the extension of Eq. (\ref{larkin}) connecting the Larkin
length $\mathcal{L}_{d}$ and the constant $\kappa^{2}$ for the uncorrelated
disorder:
\begin{equation}
\mathcal{L}_{d}=\left(  \frac{\hslash^{4}}{m^{2}\kappa^{2}}\right)  ^{
\frac{1}{4-d}} \label{larkin-d}
\end{equation}
following from the dimension consideration. The characteristic energy reads:
\begin{equation}
\mathcal{E}_{d}=\frac{\hslash^{2}}{2m\mathcal{L}_{d}^{2}}=\frac{1}{2}\left(
\frac{m^{d}\kappa^{4}}{\hslash^{2d}}\right)  ^{\frac{1}{4-d}}
\label{L-energy-d}
\end{equation}
The extension of the exponential law (\ref{exp}) for the probability to find
the deep potential well of the size not larger than $R$ is:
\begin{equation}
q\left(  R\right)  =f\exp\left[  -\left(  \frac{\mathcal{L}_{d}}{R}\right)
^{4-d}\right]  =f\exp\left[  -\left(  \frac{\left\vert E\right\vert
}{\mathcal{E}_{d}}\right)  ^{\frac{4-d}{2}}\right]  , \label{exp-d}%
\end{equation}
where the Larkin length differs from that given by definition (\ref{larkin-d})
by the factor $\left(  \Omega_{d}/2\right)  ^{\frac{1}{4-d}}$, in which
$\Omega_{d}$ is the volume of the sphere with the radius 1 in $d-$dimensional
space. As earlier $f$ is a power-like function of the ratio
${\mathcal{L}}/{R}$ at ${\mathcal{L}}/{R}\gg 1$: $f(x)\propto x^{\alpha}$
with $\alpha=1$ for $d=2,3$ and $\alpha =0$ for $d=1$ \cite{cardy}. The extension of the 
Eq. (\ref{dens-wells}) for the density of the 
fluctuation potential wells with the radius not exceeding $R$ looks in $d$
dimensions as follows:
\begin{equation}
n_{w}\left(  R\right)  =R^{-d}f\exp\left[  -\left(  \frac{\mathcal{L}_{d}}
{R}\right)  ^{4-d}\right]  \label{dens-wells-d}%
\end{equation}
and the average distance between such wells is
\begin{equation}
d\left(  R\right)  =Rf^{-1/d}\exp\left[  \frac{1}{d}\left(  \frac
{\mathcal{L}_{d}}{R}\right)  ^{4-d}\right]  \label{distance-R-d}%
\end{equation}
The tunneling amplitude between nearest wells has a characteristic value
\begin{equation}
t\left(  R\right)  =\exp\left[  -f^{-1/d}\exp\left[  \frac{1}{d}\left(
\frac{\mathcal{L}_{d}}{R}\right)  ^{4-d}\right]  \right]
\label{tunneling-R-d}%
\end{equation}
Next we consider the filling of the potential wells by the diluted Bose-gas of
the density $n=N/L^{d}$. Again, let the gas fills all the wells up to the
radius $R$. The average number of particles per well is:
\begin{equation}
\mathcal{N}\left(  R\right)  =\frac{n}{n_{w}\left(  R\right)  }=f^{-1}
nR^{d}\exp\left[  \left(  \frac{\mathcal{L}_{d}}{R}\right)  ^{4-d}\right]
\label{number -per-well-d}%
\end{equation}
and the density of particles in the well is
\begin{equation}
n_{p}\left(  R\right)  =\mathcal{N}\left(  R\right)  /\left(  \Omega_{d}%
R^{d}\right)  =\frac{n}{f\Omega_{d}}\exp\left[  \left(  \frac{\mathcal{L}_{d}
}{R}\right)  ^{4-d}\right]  \label{particle-density-in-well-d}%
\end{equation}
The repulsion energy per particle is as before $g_{d}n_{p}\left(  R\right)  $,
but the connection between the coupling constant $g$ and the scattering
amplitude $a$ depends on dimensionality of the space:%
\begin{equation}
g_{d}=\frac{\Omega_d\hslash^{2}a^{d-2}}{m} \label{g-a-d}
\end{equation}
At $d=2$ this equation is invalid, but the coupling constant $g_2$ is well
defined and has dimensionality energy$\times$length$^{2}$. Minimizing the
energy per particle in complete analogy with what was done in Section
\ref{Gas-in-box}, we find the maximal radius of the filled well as a function
of density:
\begin{equation}
R\left(  n\right)  =\frac{\mathcal{L}_{d}}{\left(  \ln\frac{n_{c}}{n}\right)
^{\frac{1}{4-d}}}, \label{R-n-d}%
\end{equation}
and the critical density is defined by the equation:
\begin{equation}
n_{c}=\frac{\Omega_{d}\mathcal{E}_{d}}{g_{d}(4-d)}=\frac{\Omega_{d}}{4\pi
\mathcal{L}_{d}^{2}a^{d-2}(4-d)} \label{n-crit-d}
\end{equation}
In 1-d Bose-gas the critical density increases with the repulsion strength in
contrast to 2- and 3-dimensional cases. The reason is that the 1-d Bose-gas
with strong repulsion (hard core) is equivalent to the Fermi-gas whose kinetic energy
increases with the size of the core. 

The ratio $n_{c}/n$ can be rewritten in equivalent form:
\begin{equation}
\frac{n_{c}}{n}=\left(  \frac{\xi_{d}}{\mathcal{L}_{d}}\right)  ^{2},
\label{nc-n-xi}
\end{equation}
where $\xi_{d}=\hslash/\sqrt{2mgn}$ is the coherence or healing length of
the Bose-gas without disorder. The condition $n\ll n_{c}$ at which the deeply
localized ground state is realized implies also $\mathcal{L}_{d}\ll\xi_{d}$ as
it could be expected. In 1d case large values of the gas parameter $na_{1}$
correspond to the almost ideal case in which kinetic energy is much larger
than the interaction. Accordingly, the gas parameter inside the fluctuation
well $n_{c}a^{d}\sim\left(  a/\mathcal{L}_{d}\right)  ^{2}$ must be large in
1d case.

We do not show other calculations analogous to those produced in Section
\ref{Gas-in-box} and present only the final results:

\noindent\textit{Number of particles in a filled well}:%
\begin{equation}
\mathcal{N}\left(  n\right)  =\Omega_{d}^{-1}\left(  \ln\frac{n_{c}}%
{n}\right)  ^{-\frac{d}{4-d}}\left(  \frac{\mathcal{L}_{d}}{a}\right)  ^{d-2}     
\label{number-in-well-d}%
\end{equation}

\noindent\textit{Distance between the filled wells}:%
\begin{equation}
d\left(  n\right)  =\mathcal{L}_{d}\left(  \frac{n_{c}}{fn}\right)
^{1/d}\left(  \ln\frac{n_{c}}{n}\right)  ^{-\frac{1}{4-d}} \label{d-n-d}%
\end{equation}

\noindent\textit{Chemical potential}:%
\begin{equation}
\mu\left(  n\right)  =-\mathcal{E}_{d}\left(  \ln\frac{n_{c}}{n}\right)
^{\frac{2}{4-d}} \label{chem-pot-d}%
\end{equation}
\textit{Tunneling amplitude between wells}:
\begin{equation}
t\left(  n\right)  =\exp\left[  -\left(  f\frac{n_{c}}{n}\right)
^{1/d}\right]  \label{t-n-d}%
\end{equation}

Considering the gas in a trap which has a shape of a disc or a cigar, one
should express the parameters of effective $d-$dimensional problem in terms of 
parameters of the 3d gas:
the  particle density $n$, 3d-scattering length $a$, Larkin length
$\mathcal{L}$, the transverse oscillator length $\ell_{\bot}$ and the
longitudinal oscillator length $\ell$. To simplify the results we consider only
the most interesting and experimentally accessible situation of a weak
confinement:
\begin{equation}
a\ll \ell_{\bot}\ll\mathcal{L} \label{inequality}
\end{equation}
Then it is straightforward to see that the effective $d-$dimensional Larkin
length $\mathcal{L}_{d}$ is related to the 3d Larkin length $\mathcal{L}$ by a
following relationship \cite{Nattermann08};
\begin{equation}
\mathcal{L}_{d}\approx\left(  \mathcal{L}\ell_{\bot}^{3-d}\right)  ^{\frac
{1}{4-d}} \label{Larkin-d-Larkin-3}
\end{equation}
Indeed, the argument in exponent of the Gaussian distribution for uncorrelated
disorder is $\sim U^{2}\Omega/2\kappa^{2}$. The volume $\Omega$ can be
approximately factorized: $\Omega\simeq\Omega_{d}R^{d}\Omega_{3-d}\ell_{\bot
}^{3-d}$. Using the standard estimate for $U=-\frac{\hslash^{2}}{mR^{2}}$, we
arrive at Eq. (\ref{exp-d}) with $\mathcal{L}_{d}$ defined by Eq.
(\ref{Larkin-d-Larkin-3}) with precision of a numerical factor. To find the
effective scattering length $a_{d}$, one should to use an obvious identity:
$ng=n_{d}g_{d}$ (both expressions are the interaction energy per particle) with 
$n_{d}=n\ell_{\bot}^{3-d}\Omega_{3-d}$ being the density in the reduced 
dimensionality space. From these two
equations it follows that:
\begin{equation}
g_{d}=\frac{g\ell_{\bot}^{d-3}}{\Omega_{3-d}} \label{g-d}%
\end{equation}
Employing Eqs. (\ref{g-a-d}) and (\ref{g-a}), we arrive at following
relationship:
\begin{equation}
a_{d}\sim\left(  a~\ell_{\bot}^{d-3}\right)  ^{\frac{1}{d-2}} \label{a-d-a}%
\end{equation}
This equation is invalid for $d=2$ and instead Eq. (\ref{g-d}) must be used.
Below we consider possible regimes similarly to what we did for 3d system.

\noindent\textbf{Weak disorder}: $\mathcal{L}_{d}\gg \ell$ or equivalently
$\mathcal{L}\gg\frac{\ell^{4-d}}{\ell_{\bot}^{3-d}}$. The random potential is
negligible. The competition of the kinetic energy, trap energy and interaction
provides two different regimes:

\noindent\textit{Weak interaction}: $g_{d}N\ll\hslash^{2}\ell^{d-2}/m$ or
equivalently $Na\ll \ell^{d-2}\ell_{\bot}^{3-d}$. In this case the longitudinal size
of the cloud $R$ coincides with $\ell$.

\noindent\textit{Strong interaction}:\textit{ }$Na\gg \ell^{d-2}\ell_{\bot}^{3-d}$.
This is the region of the Thomas-Fermi regime: $R\sim\left(  Na\ell^{4}/\ell_{\bot
}^{3-d}\right)  ^{1/\left(  d+2\right)  }$.

\noindent\textbf{Strong disorder}: $\mathcal{L}_{d}\ll \ell$ or equivalently
$\mathcal{L}\ll\frac{\ell^{4-d}}{\ell_{\bot}^{3-d}}$.

\noindent\textit{Weak interaction}: $\frac{gN}{\mathcal{L}_{d}^{d}\ell_{\bot
}^{3-d}}\ll\frac{\hslash^{2}}{m\mathcal{L}_{d}^{2}}$ or equivalently
$Na\ll\mathcal{L}^{\frac{d-2}{4-d}}\ell_{\bot}^{\frac{2\left(  3-d\right)  }
{4-d}}$. This is non-ergodic situation: all particles fall into the deepest
potential well of the typical size $R\sim\mathcal{L}_{d}/\left(  \ln\frac
{\ell}{\mathcal{L}_{d}}\right)  ^{\frac{1}{4-d}}$. For $d=2$ it happens at
$N\ll\frac{\ell_{\bot}}{a}$; for $d=1$ the non-ergodic regime exists if $\ell_{\bot
}\gg\left(  a^{3}\mathcal{L}\right)  ^{1/4}$.

\begin{figure}
\begin{center}
\includegraphics[width=0.7\linewidth]{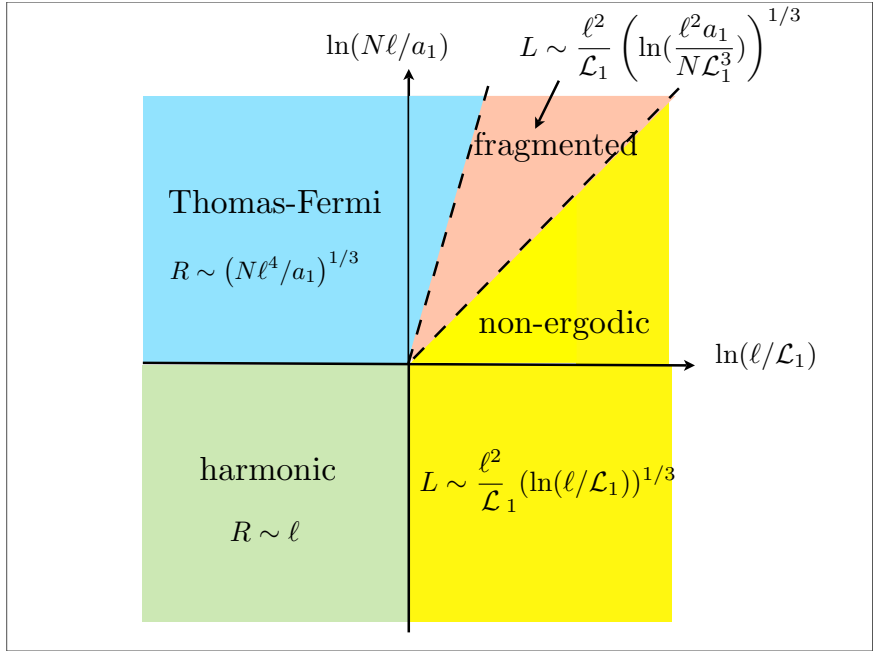}
\caption{\small{Regime diagram of atoms in a one dimensional trap: uncorrelated disorder. In reduced dimensions
$R$ denotes the longitudinal size of the single atomic cloud, $L$ denotes the longitudinal size of the fragmented state.}}
\label{Figure6}
\end{center}
\end{figure}

\noindent\textit{Moderate interaction}: $\mathcal{L}^{\frac{d-2}{4-d}}\ell_{\bot
}^{\frac{2\left(  3-d\right)  }{4-d}}\ll Na\ll \ell^{d-2}\ell_{\bot}^{3-d}$. In this
range of interaction the cloud consists of multiple fragments. They are weakly
coupled each to other provided the dimensionless parameter $\Gamma
=\ell^{2d}/\left(  Na_{d}^{d-2}\mathcal{L}_{d}^{d+2}\right)  =\ell^{2d}/\left(
Na\mathcal{L}^{\frac{d+2}{4-d}}\ell_{\bot}^{\frac{2\left(  3-d\right)  \left(
d-1\right)  }{4-d}}\right)  $ is large (note that the exponent at $\ell_{\bot}$ is zero
at $d=1,3$ and -1 at $d=2$). We remind that $\Gamma$ is equivalent
to the ratio $n_{c}/n$. A typical longitudinal size of the fragment is
$R=\mathcal{L}_{d}/\left(  \ln\Gamma\right)  ^{1/\left(  4-d\right)  }$; a
typical distance between fragments is $d=R\Gamma^{1/d}$; the tunneling
amplitude between fragments is $t\sim\exp\left(  -\Gamma^{1/d}\right)  $. The
transition from localized state to the phase-correlated superfluid state
proceeds at $\Gamma\approx1$. At $\Gamma\gg1$ the disorder is negligible and
the Thomas-Fermi approximations works. The phase diagram in one dimension is shown in Fig. \ref{Figure6}

\subsection{Strongly correlated disorder}

As in 3d case, the distribution of random potential is characterized by the
correlation length $b$ and the characteristic energy $U_{0}$. It is assumed
that $b\gg\mathcal{L}\sim\left[  \hslash^{4}/\left(  m^{2}U_{0}^{2}
b^{d}\right)  \right]  ^{\frac{1}{4-d}}$ or equivalently $b\gg B=\frac
{\hslash}{\sqrt{mU_{0}}}$. The characteristic size of a deep potential well is
equal to the correlation length $b$. The probability to find the well
containing the level with energy $E<0$ and sufficiently deep, so that
$\left\vert E\right\vert \gg U_{0}$, is given by the Gaussian exponent Eq.
(\ref{probability-corr}). The number of such wells per unit volume is
\begin{equation}
n_{wd}\left(  E\right)  =b^{-d}\exp\left(  -\frac{E^{2}}{2U_{0}^{2}}\right)
\label{n-w-d}
\end{equation}
and the distance between them $d_{d}(E)=\left[  n_{wd}\left(  E\right)
\right]  ^{-1/d}$. The gas with the average density $n_{d}$ fills the wells up
to the depth $\left\vert E\right\vert =U_{0}\sqrt{\ln\left(  n_{cd}
/n_{d}\right)  }$. The critical density is given by equation:
\begin{equation}
n_{cd}=\frac{\Omega_{d}}{d}\frac{U_{0}}{g_{d}}=\frac{\Omega_{d}}{4\pi d}%
\frac{1}{B^{2}a_{d}^{d-2}} \label{n-c-d}%
\end{equation}
The latter formula is invalid at $d=2$, but the former one, employing the
value $g_{d}$, remains correct. The particles in each well fills a
d-dimensional sphere of the radius $R=b\frac{U_{0}}{\left\vert E\right\vert
}=\frac{b}{\sqrt{\ln\left(  n_{cd}/n_{d}\right)  }}$. The transition to the
superfluid state proceeds at $n_{d}=n_{cd}$.

Proceeding to the gas in a trap, we first specify conditions which are
realistic and allow to reduce the number of possible regimes:
\begin{equation}
a\ll \ell_{\bot}\ll\mathcal{L}\ll b \label{conditions-corr-d}
\end{equation}
At these conditions $g_{d}\ell_{\bot}^{3-d}=g$; $a_{d}^{d-2}\ell_{\bot}^{3-d}=a$ and
Eq. (\ref{n-c-d}) can be rewritten as follows:
\begin{equation}
n_{cd}=\frac{\Omega_{d}}{d}\frac{U_{0}\ell_{\bot}^{3-d}}{g}=\frac{\Omega_{d}
}{4\pi d}\frac{\ell_{\bot}^{3-d}}{B^{2}a} \label{n-c-d-t}
\end{equation}
Note, that this equation is valid at $d=2$ as well. The possible regimes are
classified as follows.

\noindent1. \textbf{Weak disorder}: $U_{0}\ll\hslash\omega$ or equivalently
$B\gg \ell$.

\noindent1a. \textit{Weak interaction}: $Na\ll \ell^{d-2}\ell_{\bot}^{3-d}$.  In
this case the size of the cloud $R=\ell$.

\noindent1b. \textit{Strong interaction}: $Na\gg \ell^{d-2}\ell_{\bot}^{3-d}$. The
Thomas-Fermi range:
\begin{equation}
R=\left(  \frac{4\pi d}{\Omega_{d}\Omega_{3-d}}\frac{Na\ell^{4}}{\ell_{\bot}^{3-d}
}\right)  ^{\frac{1}{d+2}} \label{TF-d}
\end{equation}

\noindent2. \textbf{Strong disorder}: $U_{0}\gg\hslash\omega$ or $B\ll \ell$.

\noindent2a. \textit{Weak interaction}: $\frac{gN}{b^{d}\ell_{\bot}^{3-d}}\ll
U_{0}$ or equivalently $Na\ll\frac{b^{d}\ell_{\bot}^{3-d}}{B^{2}}$. Non-ergodic
situation. A typical depth of the deepest potential well is $E=-U_{0}
\sqrt{2d\ln\left(  L/b\right)  }$. A typical distance of this well from the
center of the trap is $L\sim\sqrt{U_{0}/\left(  m\omega^{2}\right)  }$
$\left(  2d\ln\frac{\ell^2}{b B}\right)  ^{1/4}=\frac{\ell^{2}}{B}\left(
2d\ln\frac{\ell^2}{b B}\right)  ^{1/4}$.

\begin{figure}
\begin{center}
\includegraphics[width=0.7\linewidth]{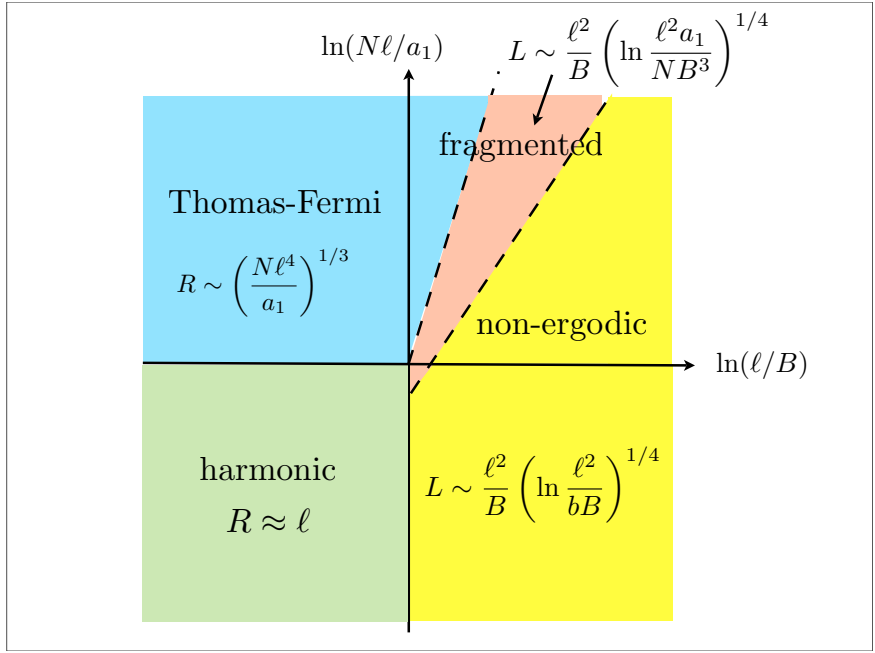}
\caption{\small{Regime diagram of atoms in a one-dimensional trap: correlated disorder. In reduced dimensions
$R$ denotes the longitudinal size of the single atomic cloud, $L$ denotes the longitudinal size of the fragmented state.}}
\label{Figure7}
\end{center}
\end{figure}

\begin{figure}
\includegraphics[width=0.7\linewidth]{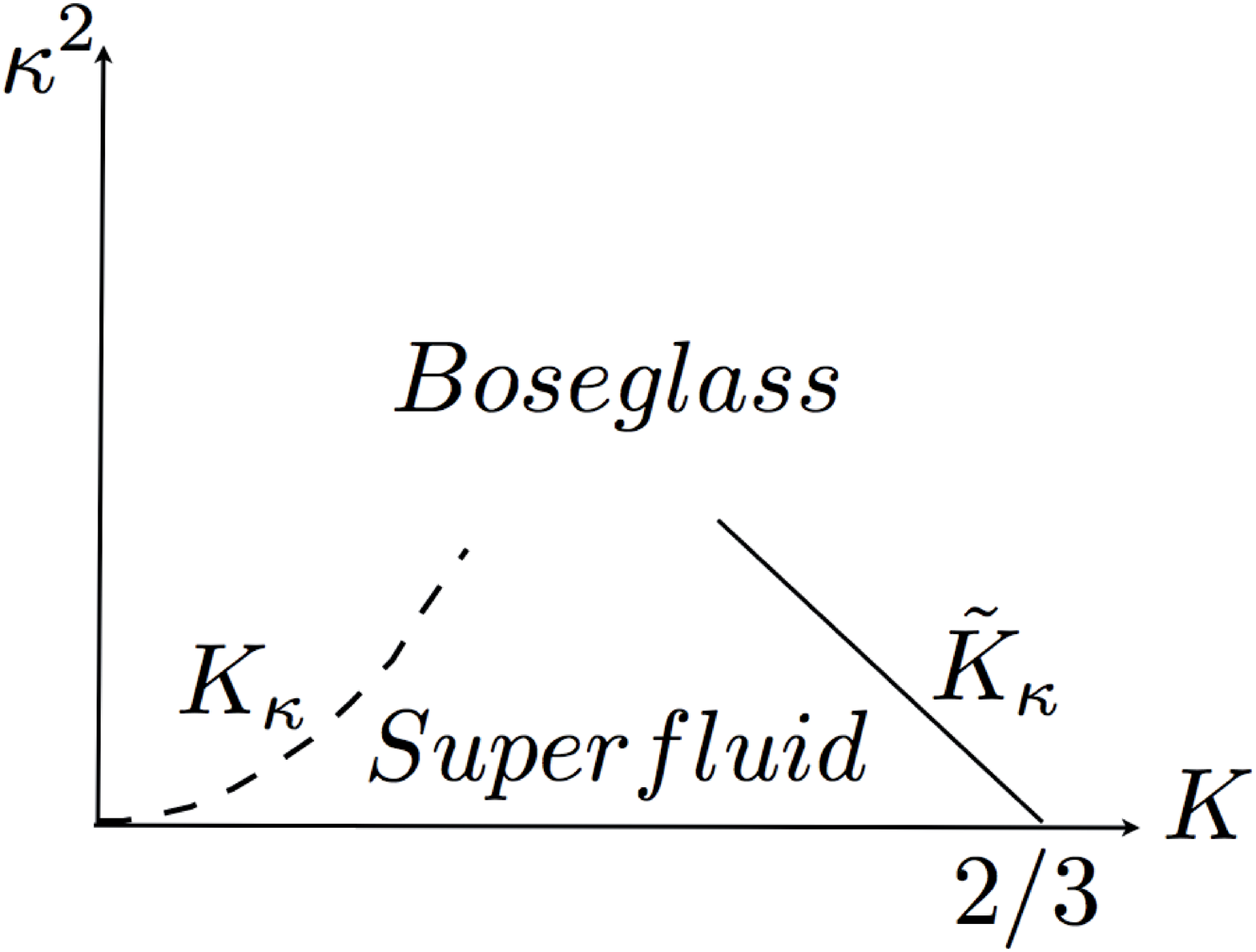}
\caption{\small{Phase diagram in one dimension. }}
\label{Figure8}
\end{figure}

\noindent2b. \textit{Intermediate interaction}: $\frac{b^{d}\ell_{\bot}^{3-d}
}{B^{2}}\ll Na\ll \ell^{d-2}\ell_{\bot}^{3-d}$. This is the range of fragmented
state. The particles in fragments are deeply localized if the parameter
$\Gamma=\ell^{2d}\ell_{\bot}^{3-d}/\left(  B^{d+2}a N\right)$ is large.         
The particles in a fragment occupy a d-dimensional sphere of the radius
$R\left( N\right)  =b/\sqrt{\ln\Gamma}$. Their typical energy is
$E=-U_{0}\sqrt{\ln\Gamma}$. A typical distance between fragments is
$d\left(  N\right)  =b\Gamma^{1/d}$. The number of particles inside each
fragment is about $b^{d}\ell_{\bot}^{3-d}/\left(  aB^{2}\right)  $. The total
size of the cloud is $\sim\left(  \ell^{2}/B\right)  \left(  \ln\Gamma
\right)  ^{1/4}$. The transition to the superfluid states proceeds at
$\Gamma\approx1$. The phase diagram in $d$-dimensions is shown in Fig. \ref{Figure7}

\section{Discussion and conclusions\label{concl}}

The theory allows prediction of many quantitative characteristics of deeply
localized states. It also gives simple formulas for the critical density
$n_{c}$ at which the quantum phase transition from the random singlet to
superfluid state proceeds. Below we discuss how these predictions can be
checked in experiments with cooled atomic gases.
Four important parameters can be controllably and independently varied in the experiment. They are: number of particles $N$;
the frequency $\omega$ or equivalently the strength of the trap; the scattering length $a$ (by
approaching one of the Feshbach resonances); the strength of disorder $U_{0}$.
Using this freedom it is feasible to pass all spectrum of regimes described
above. In particular, varying $U_{0}$ it is possible to transit from
uncorrelated to strongly correlated disorder. A simple estimate shows that, at $b\sim1\mu m$, the transition from uncorrelated to strongly correlated regime  proceeds
at frequency of disorder potential $\omega_{d}=\sqrt
{2U_{0}/mb^{2}}\sim1kHz$ which is accessible.

Simplest experiments are the measurements of the cloud size $L$ as a function of 
different variable parameters in the regime of multiple localized fragments. Our 
theory predicts that in the regime of uncorrelated disorder the size of the cloud 
is inversely proportional to the frequency of the harmonic trap $\omega$ 
and proportional to the square of the disorder strength. It also predicts a weak 
dependence of the size on the number of particles $\sim{\ln N}$. In the
case of strongly correlated disorder the size of the cloud is again inverse proportional 
to $\omega$, but the dependence on the disorder strength is much weaker
$L\propto U_{0}^{1/2}$. The dependence on $N$ also is weaker than in the uncorrelated regime: 
$L\propto\left(  \ln N\right)  ^{1/4}$.
It would be important to observe a crossover from non-ergodic state with
one or few fragments to the ergodic state with many fragments and check that
it happens at $N=\mathcal{L}/3a$ for uncorrelated disorder and at
$N=\frac{b^3}{3aB^2}$ for strongly correlated disorder (see modifications of these results for
$d=1,2$ in Section 7). Counting the number of fragments, one can find the
number of particles per fragment and compare it with theoretical prediction.
Direct measurements of the size of fragments and distances between them are
much desirable since they give an important information on the balance
between the inter-particle repulsion and their interaction to random potential.

Another feasible experiment is the time-of-flight spectroscopy after switching
off both the trap and the random potential. In this experiment the
distribution of particles over momenta (velocities) is measured. The
distribution of momentum has a finite width $\Delta p$. According to the
uncertainty principle the value $\hslash/\Delta p$ can be treated as the
average size of the fragment $R=\mathcal{L}/\ln\Gamma$ for the uncorrelated
disorder (see the definition of $\Gamma$ in Eq. (\ref{Gamma})). The fastest is
the dependence of this value on the strength of disorder:\ $\Delta p\propto
U_{0}^{2}$. Others dependencies are logarithmic. For example $\frac{d\Delta p}
{d\ln\omega}=-\frac{3\hslash}{\mathcal{L}}=3\frac{d\Delta p}{d\ln N}$.
Installing a counter close to the trap at a distance comparable to the size of
the trap would allow to register the oscillations of the particle flux due to
discrete character of the fragmented state. This is another opportunity to
find the distances between fragments. In the case of strongly correlated
disorder the width of the distribution is determined by Eq.
(\ref{momentum-corr}), in which the ratio $n_{c}/n$ must be substituted by
$\Gamma$ defined by Eq. (\ref{Gamma-corr}). In this case $\Delta p$ depends
linearly on $\omega$. Other dependencies are much weaker. Nevertheless, the
prediction that the ratio $\Delta p/\omega$ depends on one dimensionless
parameter $\Gamma$ can be experimentally checked.

The transition between localized and delocalized coherent state in the random
potential was confirmed in several experiments (see Introduction). We propose
to make more detailed measurement of the transition manifold and check our
predictions formulated in Eqs. (\ref{n-crit}, \ref{crit-density-corr}).

An important question is whether the relaxation to the ground state can be
reached during a reasonable time interval compatible with the time of
experiment. We analyze this question for the uncorrelated or weakly correlated
disorder. In this case the relaxation time due to tunneling can be estimated
as $\tau=2\pi\omega_{n}^{-1}t^{-1}$, where $\omega_{n}\sim\frac{\mathcal{E}
}{\hslash}\left(  \ln\Gamma\right)  ^{2}$ is the characteristic frequency of
the optimal potential well and $t\sim\exp\left[  -\Gamma^{1/3}\right]  $ is
the tunneling coefficient (see Eq. (\ref{tunneling-n})). For numerical
estimates we accept $\Gamma\sim125$, $\ell\sim10\mu m$, $b\sim\mathcal{L}
\sim1\mu m$, $a\sim0.01\mu m$, $N\simeq27,000$. Then $t^{-1}=148$ and
$\tau\sim0.06s$. The Larkin length can be increased by decreasing the
amplitude of the random potential ($\mathcal{L}\propto U_{0}^{-2}$).
Simultaneously at fixed values $N$, $\ell$ and $a$ the value $\Gamma$
decreases as $\mathcal{L}^{-5}$. This example shows that the equilibrium is
accessible, though it may be difficult to reach large ratio $\mathcal{L}/b$.
In experimentally important case of 1d (cigar-like) cloud the relaxation
time is shorter since it is proportional to $\exp (-\Gamma_1)$ with 
$\Gamma_1=\frac{\ell^2}{3Na\mathcal{L}}$. At $\ell =25\mu m$, $\mathcal{L}=1\mu m$ 
and $a=0.01 \mu m$ the ratio $\Gamma_1$ reaches value 1 at $N\sim 85,000$.

In conclusion, we presented a simple physical picture of deeply localized
states of the Bose gas in a random potential. We demonstrated that the
particles eventually fill the deep potential wells formed by fluctuations of
the random potential and by their self-consistent field. Based on this idea
the geometrical and physical properties of these states, such as the size of
the clusters, the distances and the tunneling amplitudes between them, the
size of the whole cloud formed by the gas in a harmonic trap were calculated
with precision of numerical coefficients. We
discovered that the physical properties of these states depend significantly
on correlations of the random potential. It occurs that the ground state of
the system can be either almost homogeneous and coherent (superfluid) if the
disorder is weak enough, or fragmented and strongly localized. In particular,
if the disorder is much stronger than the repulsion between particles, the
system transits  into a  non-ergodic state, whose properties even in the equilibrium
strongly depend on the specific sample. At growing number of particles the
system occurs in an ergodic, but strongly localized ground state consisting of
multiple particle clusters populating deep fluctuation wells. At the  number of particles increasing the tunneling between different potential
wells increases leading to the phase correlation and finally to the quantum
phase transition to the coherent (superfluid) state. We have found simple
expression for the gas density at this transition in the gas confined by a big
box and equation for the phase transition manifold in the gas confined by the
harmonic trap. It was shown that in any dimension $d<4$ the repulsive interaction causes delocalization and quantum phase transition to the superfluid state at a critical density. Thus, the interaction overcomes the Anderson localization of all single-particle states in 1 and 2 dimensions. In 3 dimensions the single particle levels which are filled at critical density have energy still below the mobility threshold.\bigskip

\section*{Acknowledgements}
The authors acknowledge a helpful support from the DFG through NA222/5-2 and
SFB 680/D4(TN) as well as from the DOE under the grant DE-FG02-06ER46278.

\section*{Appendix: The superfluid-insulator transition in one dimension}

In this appendix we discuss the relation of our approach to the theory of Giamarchi and Schulz  \cite{Giamarchi}.
It is convenient to state first the relation between   the mass  $m$, the boson density $n$,
the scattering length $a$, the parameters $K$ and $\kappa$ of the field theory of 1d bosons \cite{Haldane, Giamarchi}. We
omit all subscripts $_1$ which are excessive for one dimension. The phonon velocity $v$ is then given by
\begin{equation}
v =\left(\frac{n}{m\varkappa}\right)^{1/2}
\end{equation}
where $\varkappa$ denotes the compressibilty of the bosons
\begin{equation}
\varkappa=\frac{\partial n}{\partial \mu}=\frac{1}{\pi\hbar vK}.
\end{equation}
The Haldane inverse compressibility $\varkappa_H$
\cite{Haldane} is related to our standard definition by $\varkappa=n^2/\varkappa_H$.
Indeed,  according to our definition the elastic energy per unit length is 
$E\approx\frac{1}{2\varkappa}{\delta n}^2$
which implies
\begin{equation}
K=\frac{1}{\pi\hbar} \left(\frac{m}{n\varkappa}\right)^{1/2}.
\end{equation}
The Boson coupling constant $K$ is inverse to that of fermions.
These relations agree with those of  \cite{Fisher} taking in account that in \cite{Fisher} the density is dimensionless and $\hbar=1$. To get the relation between
$a$ and $K$ we quote Haldane \cite{Haldane} referring to earlier work of Lieb \cite{Lieb}. Using the dimensionless parameter
$\gamma={1}/({\pi^2na})$
(we absorbed a factor $1/\pi^2$ in the definition of $\gamma$), $K$ is related to $\gamma$ by
\begin{eqnarray}
K&\approx&\gamma^{1/2}\left(1- \frac{1}{2}\gamma^{1/2}\right),
\quad \gamma \ll1 \\
K&\approx&\left(1-\frac{8}{\pi^2\gamma}\right)^{1/2},   \quad \gamma \gg 1.
\end{eqnarray}
with a smooth interpolation in between.
Thus $K$ changes continuously from $0$ (for $a\to \infty$ corresponding to no
interaction) to $K=1$ (for infinitely strong interaction).
Since for $K=1$ bosonic and fermionic description coincide,
an infinitely strong repulsive interaction corresponds to free fermions \cite{Egger}.
Let us formulate the condition of superfluidity in terms of coupling constant $K$.  
The condition of superfluidity for 1d \emph{disordered} system $n>n_c\approx a /{\cal L}^2$
(see Eq. (\ref{eq:n_c(d)})) can be rewritten as
\begin{equation}
 K>K_{\kappa}=\frac{1}{\pi n}\left(\frac{m\kappa}{\hbar^2}\right)^{2/3}\equiv \frac{1}{\pi n {\cal L}}.
\end{equation}
 where $\kappa$ denotes the disorder strength.                                
 We will show that $K\approx K_{\kappa}$ remains the phase boundary between the superfluid and
 Bose glass phases also if quantum fluctuations are taken into account.                                    

Let us study how quantum fluctuations 
renormalize 
the disorder strength $\kappa$. In lowest order in ${\kappa}$  the disorder strength 
is renormalized by the Debye-Waller factor:
\begin{equation}
\kappa_{\mathrm{eff}}\approx \kappa
e^{-2\langle \varphi^2\rangle}.
\end{equation}
Here
\begin{equation}
2\langle\varphi^2\rangle=\frac{2\pi}{K}\int_{{\cal L}^{-1}}^{\xi^{-1}}
\frac{d^2q}{(2\pi)^2}\frac{1}{q^2}\approx\frac{1}{K}\ln\left( {{\cal L}/\xi}\right)\approx
\frac{1}{K}\ln \left(\frac{K}{K_{\kappa}} \right)\end{equation}
where we have used the relation for small $K$. 
Thus, for $K<K_{\kappa}$ there are no fluctuation corrections, and our condition (\ref{eq:n_c(d)})
tells us that there is no superfluidity. In the opposite case $K>K_{\kappa}$ we find (by making
the calculation self-consistent) for max$(K_{\kappa},2/3)<K$
\begin{equation}
\kappa_{\mathrm{eff}}\approx \kappa\left(\frac {K_{\kappa}}{K}\right )^{\frac{1}{K-2/3}},\quad \end{equation}
Approaching the critical value $K=2/3$ the effective disorder strength is renormalized to zero   for
 $K_{\kappa}<K<2/3$. In the opposite case there is no renormalization and the disorder is relevant.
 From this considerations it is clear that the transition happening at $K=2/3$ occurs only in one dimension. There is another transition line $K=K_{\kappa}$ which 
exists in higher dimensions.
 In the next order in $\kappa$, the condition  $K=2/3$ changes to $K=\tilde K_{\kappa}$ with
 $\tilde K_{\kappa=0}=2/3$  \cite{Giamarchi}. The transition at $K=\tilde K_{\kappa}$ is of the Kosterlitz-Thouless type. 
Even in a \emph{pure} system superfluidity persists only when
the healing length $\xi$ is smaller than the system size $L$, i.e. for $\xi\sim (a/n)^{1/2}< L$. For small $K$ this condition can be written as $\pi nK>1/ L$.
Thus,  for $a\to\infty $ the superfluidity persists  only if

\begin{equation} K> K_L=\frac{1}{\pi n
L}.
\end{equation}


\begin{thebibliography}{99}
\bibitem{d-w} The value which can be defined exactly is the density $n_w(E)$ of states
with the energy less than $E$: $n_w(E)=\int_{-\infty}^E \nu (E) dE$. For energies corresponding
to deeply localized states ($E<0$ and $\mid E\mid\gg \cal{E}$) this value coincides with $n_w(R)$
at $R=\hbar /\sqrt{2m\mid E \mid}$.
\bibitem{regime} The regime diagram is the analogue of the phase diagram. 
The difference is that the sharp phase transitions are realized only in
infinite systems, whereas in finite systems a more or less smooth crossovers
between different regimes take place.
\bibitem {quasi-1d}Quasi-one-dimensional system in experiments with cooled
gases was realized as a narrow cylinder (or ellipsoid) of the length $\ell$
and transverse extension $\ell_{\bot}$, with $\ell\gg\ell_{\bot}\gg a$. In
this case we get $a_{1}\sim\ell_{\bot}^{2}/a$ and $n_{1}=n\ell_{\bot}^{2}$,
respectively. The dimensionless interaction parameter can be rewritten as
$\left(  a_{1}n_{1}\right)  ^{-1}=an^{1/3}\left(  n^{1/3}\ell_{\bot}\right)
^{-4}$. Thus, if the gas parameter $an^{1/3}$ is small and the radius of the
cylinder is larger than the mean particle distance, we get $\left(  a_{1}
n_{1}\right)  ^{-1}\ll1$. In this region the coupling constant $K$ satisfy a
following relation $K\sim\sqrt{an^{1/3}}\left(  n^{1/3}\ell_{\bot}\right)
^{-2}\ll1$.
\end{thebibliography}
\end{document}